\documentclass[12pt]{article}

\usepackage{jfe_preambule}
\usepackage{booktabs}
\usepackage{color}
\usepackage{colortbl}
\usepackage{multicol}
\usepackage{breakcites}

\newcommand{\Cov}{\mathbb{C}\mathrm{ov}}
\newcommand{\Var}{\mathbb{V}\mathrm{ar}}
\newcommand{\Prob}{\mathbb{P}}

\newtheorem{remark}{Remark}[section]
\newtheorem{definition}{Definition}[section]

\begin{document}

\setlist{noitemsep}  

\title{Testing Sharpe ratio: luck or skill?}

\author{Eric Benhamou 
\thanks{A.I. SQUARE CONNECT, 35 Boulevard d'Inkermann 92200 Neuilly sur Seine, France}  
\textsuperscript{,} 
\thanks{LAMSADE, Université Paris Dauphine, Place du Maréchal de Lattre de Tassigny, 75016 Paris, France} 
,
David Saltiel
\textsuperscript{* ,} 
\thanks{Universite du Littoral Cote d’Opale, LISIC, France} 
,
Beatrice Guez
\textsuperscript{*}
,
Nicolas Paris
\textsuperscript{*}
}

\date{}              

\singlespacing

\maketitle

\vspace{-.2in}
\begin{abstract}
\noindent Sharpe ratio (sometimes also referred to as information ratio) is widely used in asset management to compare and benchmark funds and asset managers. It computes the ratio of the (excess) net return over the strategy standard deviation. However, the elements to compute the Sharpe ratio, namely, the expected returns and the volatilities are unknown numbers and need to be estimated statistically. This means that the Sharpe ratio used by funds is likely to be error prone because of statistical estimation errors. In this paper, we provide various tests to measure the quality of the Sharpe ratios. By quality, we are aiming at measuring whether a manager was indeed lucky of skillful. The test assesses this through the statistical significance of the Sharpe ratio. We not only look at the traditional Sharpe ratio but also compute a modified Sharpe insensitive to used Capital. We provide various statistical tests that can be used to precisely quantify the fact that the Sharpe is statistically significant. We illustrate in particular the number of trades for a given Sharpe level that provides statistical significance as well as the impact of auto-correlation by providing reference tables that provides the minimum required Sharpe ratio for a given time period and correlation. We also provide for a Sharpe ratio of 0.5, 1.0, 1.5 and 2.0 the skill percentage given the auto-correlation level.
\end{abstract}

\medskip



\vspace{0.5cm}

\medskip

\noindent \textit{Acknowledgement}:  The authors would like to thank Francois Bertrand for useful remarks and comments.

\clearpage

\onehalfspacing
\setcounter{footnote}{0}
\renewcommand{\thefootnote}{\arabic{footnote}}


\section{Introduction}
When an investor faces choices to invest in various funds, it is quite common for her/him to compare their Sharpe ratio. Sharpe ratios are used to rank funds because it is a single number that capture both performance and risk. As this eponymous ratio established by \cite{Sharpe_1966} is simply the ratio of the (excess) net return over the strategy standard deviation, it measures performance for a given level of volatility, which acts a proxy for the risk of the strategy. Sharpe ratio is sometimes also called information although the latter semantic alludes to a benchmark and the volatility is computed with respect to a benchmark, leading to the concept of tracking error rather than standard deviation. In the particular case of a zero percent benchmark and a zero percent risk free rate, both Sharpe and information ratio are rigorously the same. The elements that are used in the Sharpe ratio computation, namely, the returns and the volatilities are unknown numbers and need to be estimated statistically. Hence, Sharpe ratio used by investors could be error prone because of statistical estimation error. The investor face a dilemma when looking at the Sharpe ratio of potential investment. Is the printed Sharpe ratio provided by a funds real or just the result of chance? Obviously, the longer the period of the track, the more accurate the Sharpe should be. Also intuitively, the higher the Sharpe, the more chance there is that the asset manager has some real skills. The crucial question is really how to statistically measure this intuition. More precisely, we are interested in knowing if a Sharpe ratio is statistically significant. 

\subsection{Related Works}
There exists a very prolific literature on Sharpe ratio as this ratio has established itself as a benchmark for measuring performance given risk. This literature targets mostly discussing the pros and const of Sharpe ratio (\cite{PilotteSterbenz_2006},  \cite{Sharpe_1998}, \cite{Nielsen_Vassalou_2004}) leading to other performance ratios like Treynor ratio (see \cite{Treynor_FBlack_1973}), but also Calmar (see \cite{Young_1991}), Sterling (see \cite{McCafferty_2003}) or Burke ratio (see \cite{Burke_1994}). 
Other authors have also tried to extend Sharpe ratio by providing additional constraints to the Sharpe as in \cite{Bertrand_2008}, or \cite{Darolles_2012} or to use option implied volatility and skewness as in \cite{DeMiguel_Plyakha_Uppal_Vilkov_2013} or the interesting approach by \cite{Challet_2017} to compute Sharpe ratio through total drawdown duration. Last but not least, there have been also numerous empirical work on Sharpe ratio as for the most recent ones  in \cite{Giannotti_Mattarocci_2013}, \cite{AndersonBianchi_2014}.

When it comes to understand better the Sharpe ratio, the literature is more sparse. Among the few attempts to study the statistical property of the Sharpe ratio, \cite{Lo_2002} is often quoted as the founding reference. Using standard asymptotic theory under several sets of assumptions (independent normally distributed - and identically distributed returns), it provides the asymptotic distribution of the Sharpe ratio. This is interesting as it provides intuition of potential bias and correction to apply to get an unbiased estimator. \cite{Mertens_2002} and later \cite{Christie_2005} derived the asymptotic distribution of the Sharpe ratio under relaxed assumptions of stationarity and ergodicity. \cite{Opdyke_2007} interestingly showed that the
derivation provided by \cite{Christie_2005} under the non-IID returns condition was in fact identical to the one provided by \cite{Mertens_2002}. \cite{Liu_2012} and \cite{Qi_2018} improved approximation accuracy to order $O(n^{-3/2})$ . \cite{Schmid_2009} derives explicit formulas for the estimation error of Sharpe’s ratio for general correlation structures of the excess
returns. \cite{Unhapipat_2016} tackled the statistical inferences on the Sharpe ratio based on small samples while the works of \cite{Ledoit_2008}, \cite{Christie_2005}, \cite{Schmid_2009}, \cite{JobsonKorbie_1981}, \cite{Opdyke_2007} extend the Sharpe ratio analysis to the case of two investment strategies comparison. \cite{Pav_2015_a,Pav_2015_b,Pav_2016} created standard packages in R to be able to perform various tests about the Sharpe as well as presented the connection of the Sharpe to the Student distribution. Lately, \cite{Riondato_2018} provided a nice overview of most of these methods while \cite{Benhamou_2018} gave various computations for the Sharpe ratio and its tight connection to the Student distribution.

\subsection{Contributions}
The main contributions of this paper are the following:
\begin{itemize}
\item surveying and discussing impact of time aggregation as Sharpe ratio are presented on annual basis but computed on either daily or monthly data
\item surveying and presenting Sharpe ratio distributions and their corresponding statistical test
\item providing various tables for Sharpe ratio significance
\item last but not least, providing some hint about the number of trades required for having Sharpe ratio statistical significance given auto correlation.
\end{itemize}

\section{Definition}
\subsection{Standard definition}
In this section, we present the traditional definition of the Sharpe ratio. The Sharpe ratio is simply defined as the ratio of expected excess return ($ \mathbb{E}[R] - R_f$) over the risk free rate $R_f$ to its standard deviation, $\sigma$:

\begin{equation}
SR = \frac{ \mathbb{E}[R]- R_f}{\sigma} \label{eq:SR}
\end{equation}

In equation \eqref{eq:SR}, under the assumptions that returns are i.i.d., it is easy to see that the Sharpe ratio is simply the t-statistic divided by $\sqrt n$. 
Equally, $\sqrt n SR $ follows a centered or not Student distribution under the explicit assumption that the returns are normally distributed according to a normal distribution with mean $\mathbb{E}[R]$, variance $\sigma^2$ and are i.i.d.  Because $ \mathbb{E}[R]$ and $\sigma$ are unobservable, they are estimated using historical data as the population moments of the returns' distribution.
Hence, given historical returns $(R_1, R_2,..., R_n)$, the standard estimator for the Sharpe ratio (SR) is given by
\begin{equation}
SR = \frac{\hat{ \bar R} - R_f}{\hat \sigma}
\end{equation}

where 
\begin{equation}
\hat{ \bar R } =\frac{\sum_{i=1}^n R_i }{n}
\end{equation}

\begin{equation}
\hat \sigma  =  \sqrt{\frac { \sum_{i=1}^n \left( R_i - \hat{ \bar R} \right)^2 }{ n-1}} \quad  \text{or equivalently} \quad  \hat \sigma  =   \sqrt{\frac { \sum_{i=1}^n  (R_i ^2 - \hat{ \bar R}^2) }{ n-1}}
\end{equation}
Pre-computing $\hat{ \bar R }$ avoids a double summation in the definition of the volatility estimator. The $n-1$ divisor is for the estimator to be unbiased. 
Hence, the formula for SR in terms of the individual returns $\left( R_i \right)_{i=1..n}$ is given by:
\begin{equation}
SR =   \frac{\sqrt{n-1} \sum_{i=1}^n \left( R_i  - R_f \right) }{  n \sqrt{\sum_{i=1}^n \left( R_i - \hat{ \bar R} \right)^2  } } 
\end{equation}

\noindent As the non centered Student statistics writes as
$$
t = \frac{\hat{ \bar R} - R_f}{ \hat \sigma / \sqrt{n}} = \sqrt n SR
$$
It is straightforward to see that $\sqrt n SR $ follows a non-central t-distribution with degree of freedom $n-1$. 
The non-centrality parameter $\eta$ is given by 
\begin{equation}
\eta  = \sqrt n  \; \frac{\mathbb{E}[R] - R_f}{\hat{\sigma}} 
\end{equation}
 
This result has been first mentioned in \cite{Miller_1978}, largely then emphasized in \cite{Pav_2016}, \cite{Riondato_2018} and \cite{Benhamou_2018}. Accurate estimation of the corresponding bias and variance are detailed in the case of non iid returns in \cite{Bao_2006} and \cite{Bao_2009}.

\subsection{Capital dimensionless Sharpe ratios}
One of the main drawback of the traditional Sharpe ratio definition is the dependence on returns. This dependence on returns means consequently for an established track record that we are sensitive to the allocated capital as it defined the corresponding returns. If we want to have a definition that does not depend on our used capital, we can compute the Sharpe ratio not any more on returns but on the generated PnL of our track record.

\begin{definition}
Let us denote daily (resp. monthly) PnL: $(P_i^d)_{i=1,.., N^d}$ (resp. $(P_i^m)_{i=1,..,N^m}$) with $N^d$ (resp. $N^m$) the number of daily (resp. monthly) trades.\\
The 'annualized' \textit{daily capital dimensionless} Sharpe ratio $SR_{cd}^d$ is defined on daily PnL numbers ($P_i^d$):
$$
SR_{cd}^d =\sqrt{252} \,\,\,  \frac{ \frac{\sum_{i=1}^{N^d} P_i^d}{N^d}}{ \sqrt{\frac{\sum_{i=1}^{N^d} (P_i^d- \bar{P^d})^2}{{N^d}-1}}} 
$$
where $\bar{P^d} = \frac{\sum_{i=1}^{N^d} P_i^d }{N^d}$ is the average daily PnL. The $\sqrt{252}$ factor is justified in the next section by the square root rule of thumb.
for monthly periods, the 'annualized' \textit{monthly capital dimensionless} Sharpe ratio $SR_{cd}^m$ is defined on monthly PnL numbers ($P_i^m$):
$$
SR_{cd}^m  =\sqrt{12} \,\,\,  \frac{ \frac{\sum_{i=1}^{N^m} P_i^m }{N^m}}{ \sqrt{\frac{\sum_{i=1}^{N^m} (P_i^m- \bar{P^m})^2}{{N^m}-1}}} 
$$
where $\bar{P^m} = \frac{\sum_{i=1}^{N^m} P_i^m }{{N^m}}$ is the average monthly PnL. Likewise, the $\sqrt{12}$ factor is justified in the next section.
\end{definition}

\section{Time and Trade aggregation}\label{sec:time_aggregation}
\subsection{Square root rule of thumb}
Before reviewing distributions for Sharpe ratio, it is critical to discuss an important practical 
aspect of the Sharpe ratio: the Sharpe ratio is not independent from the time period $T$ 
over which excess returns $(R_i)_{i=1,\ldots,n}$ are 
computed (see \cite{Lo_2002}, or \cite{Riondato_2018}). More specifically, a Sharpe
 ratio computed over monthly returns cannot be compared directly with a Sharpe ratio 
computed over yearly return. The rule of thumb that is commonly used to multiply a $\tau$ 
period Sharpe ratio by the square root of the number of period in a year is inaccurate 
and misleading. This rule of thumb for instance gives that to infer the annual Sharpe ratio 
from daily return, one should multiply the daily Sharpe by $\sqrt{252}$ assuming $252$ data
 points per year, and by $\sqrt{12}$ for monthly Sharpe. When returns are not observed 
every day (case of intraday fund that do not necessarily trade every day), this rule of thumb
 can be extended as follows. Assume that the Sharpe ratio is computed over a fraction 
of year $yf$ and that for this period we have $d$ returns. The rescaling factor should 
be $\sqrt{d / yf}$ This more precisely practical rule allows to adjust for specific trading 
calendar that do not necessarily have 252 trading days per year.

As explained in \cite{Lo_2002}, or \cite{Riondato_2018}, this rule of thumb takes its logic and justification from i.i.d assumptions. Let us denote by $R_t(q)$  the $q$ period return:
\begin{equation}
R_t(q) \equiv R_t + R_{t-1} + \ldots+ R_{t-q+1} 
\end{equation}
where in our definition, we have ignored the effects of compounding for computational efficiency\footnote{Of course, the exact expression for compounding returns is
$ R_t(q) \equiv \prod_{i=0}^{q-1}( 1 + R_{t-j} )-1$. But the sake of clarity, we can ignore the compounding effect in our section as this will be second order effect. Equally, we could use log or continuously compounded returns defined as $\log(P_t/P_{t-1})$ in which case, our definition would be exact }. 
We have that under the assumption of i.i.d. returns whose distribution is given by $R_1$, for a period of length $T = k q$, 
$$
\Var(R_T) = \Var\left(\sum_{i=1}^T R_{i}\right) = k q \Var(R_1) 
$$
while 
$$
\Var(R_t(q))= \Var\left(\sum_{i=1}^{q} R_{t+1-i}\right) = q \Var(R_1) 
$$
which justifies the rule of thumb
\begin{equation}\label{eq:compounding}
\sqrt{\Var(R_T) } = \sqrt{q} \Var(R_t(q))
\end{equation}

The latter rule hence states that
\begin{equation}\label{eq:diff_period}
SR_{Annual} =  \sqrt{12} \,\,\, SR_{Monthly} = \sqrt{252} \,\,\, SR_{Daily}
\end{equation}

However, in the presence of auto correlation, this rule of thumb is wrong and has been explicitly emphasized in \cite{Lo_2002}. This is the subject of the following section

\subsection{AR(1) assumptions}
The i.i.d. normal assumption for the return is far from being verified in practice. A more realistic set-up is to assume that the returns follow an AR(1) process defined as follows:
\begin{equation}\label{AR_assumptions}
\left\{ {
\begin{array}{l l l l }
R_t 			&  = & \mu + \epsilon_t 				& \quad t \geq 1 ; \\
\epsilon_t 	& = & \rho \epsilon_{t-1} + \sigma v_t 	& \quad t \geq 2 ; 
\end{array} } \right.
\end{equation}

where $v_t$ is an independent white noise processes (i.i.d. variables with zero mean and unit constant variance). To assume a stationary process, we impose
\begin{equation}
\lvert {\rho} \rvert  \leq 1
\end{equation}

It is easy to check that equation \ref{AR_assumptions} is equivalent to
\begin{equation}
\begin{array}{l l l l }
R_t 			&  = & \mu + \rho ( R_{t-1} - \mu )+ \sigma v_t 	& \quad t \geq 2 ; 
\end{array}
\end{equation}

We can also easily check that the variance and covariance of the returns are given by
\begin{equation}\label{moment2}
\begin{array}{l l l l }
V(R_t) & = & \frac {\sigma^2} {1-\rho^2} 								\; \;\;\;\; \text{for} \;  t \geq 1  \\
Cov(R_t, R_u ) & = & \frac {\sigma^2 \rho^{ \lvert {t -u} \rvert  }} {1-\rho^2} 	\; \;\;\;\; \text{for} \;  t,u \geq 1 
\end{array}
\end{equation}

Both expressions in \ref{moment2} are independent of time $t$ and the covariance only depends on $\lvert {t -u} \rvert$ implying that $R_t$ is a stationary process. If we now look at the empirical SR under these assumptions, it should converge to
\begin{equation}
\frac{ \mathbb{E}[{R_t}] - R_f } { \sqrt {var(R_t)}} = \frac{ \mu - R_f}{\sqrt { \frac{ \sigma^2 }{ 1- \rho^2}}}
\end{equation}

\subsubsection{Impact of sub-sampling}
The impact of sub-sampling is considerable for Sharpe ratio as stated by the following proposition.

\begin{proposition}\label{prop:1}
The ratio between the $q$ period returns $SR(q)$ defined by
\begin{eqnarray}
SR(q) & = & \frac{ \mathbb{E}[ R_t(q)  ] - R_{f}}{ \sqrt{ Var[R_t(q)] }   }
\end{eqnarray}

and the annual (regular) SR is the following:
\begin{equation}\label{SRq_eq1}
\frac{SR(q)}{SR} = \frac{q \sigma_{\infty} }{ \sqrt{ \sum_{i=0}^{q-1} \sigma^2_{t-i} + 2  \sum_{k=1}^{q-1} \sum_{i = 0 }^{q-1-k } \rho_{t-i, t-i-k} \sigma_{t-i} \sigma_{t-i-k} } }
\end{equation}
If the return process is stationary with a constant variance  $\sigma^2 =  Var[R_t] =\sigma_{\infty}^2 $ and stationary correlation denoted by $\rho_{v-u} = Corr( R_{u},R_{v})$, this relationship simplifies to
\begin{equation}\label{SRq_eq2}
\frac{SR(q)}{SR}= \sqrt{ \frac{ q} { 1 + 2  \sum_{k=1}^{q-1} (q-k) \rho_{k} } }
\end{equation}
If in addition, the returns follow an AR(1) process $\rho_k=\rho^k$, equation \ref{SRq_eq2} becomes
\begin{equation}
\frac{SR(q)}{SR} = \sqrt{ \frac{ q} { 1 + \frac{ 2 \rho }{ 1-\rho} \left( 1 - \frac{ 1 - \rho^q}{q (1-\rho)} \right) } } 
\end{equation}

Denoting $\delta = \sqrt{ 1 + \frac{ 2 \rho }{ 1-\rho} \left( 1 - \frac{ 1 - \rho^q}{q (1-\rho)} \right)}$, we have that: 
\begin{equation}\label{SRq_eq3}
\frac{SR(q)}{SR} = \frac{ \sqrt{q} }{\delta}
\end{equation}

If the returns are non correlated ($\rho_k=0$), equation \ref{SRq_eq3} becomes
\begin{equation}\label{SRq_eq4}
\frac{SR(q)}{SR} = \sqrt{ q}
\end{equation}
The last equation is the so called square root rule that states that the annual Sharpe is equal to $\sqrt{12}$ the monthly Sharpe.
\end{proposition}

\begin{proof}
Given in \ref{proof:1}
\end{proof}

\begin{remark}
Combining equations \eqref{SRq_eq2} and \eqref{SRq_eq4}, we obtain that there is a relationship between the real Sharpe $SR(q)$ and the so called square root Sharpe 
$SR_{\mathrm{SquareRoot}} = \sqrt{ q} SR$:
\begin{equation}
SR(q) = \frac{ \sqrt{ q} SR } { \delta} = \frac{ SR_{\mathrm{SquareRoot}}} {\delta}
\end{equation}
Hence, if $\delta$ is larger than one, the real Sharpe is underestimated by the square root rule.
We obtain a major result summarized by the proposition below
\end{remark}

\begin{proposition}
If first order correlation $\rho$ is positive, the real Sharpe is underestimated by the square root rule.
If first order correlation $\rho$ is negative, the real Sharpe is overestimated by the square root rule.
\end{proposition}

\begin{remark}
The role of auto correlation in Sharpe computation has first been found by \cite{Lo_2002}.
\end{remark}

\section{Distributions for Sharpe ratio}\label{sec:distributions}
In practice, returns are not i.i.d. and suffer from auto-correlation and heteroscedasticity. Adapting the result of the previous section and inspired by 
\cite{Ledoit_2008}, we can compute the \textit{Studentized} Sharpe ratio as explained by the proposition below

\begin{proposition}\label{prop1} \textbf{Student Distribution} - 
For returns that are i.i.d. and normally distributed, we have that 
\begin{equation}\label{eq:Studentized}
\sqrt{n} \,\, SR  
\end{equation}
follows a Student distribution. In the more general (and realistic) case (as described in \cite{Benhamou_2018}), where returns follow an AR(1) process with first order auto-correlation denoted by $\rho$, and computing the Sharpe ratio on a period frequency with $q$ data points per year, the \textit{studentized} Sharpe ratio given by
\begin{equation}\label{eq:Studentized2}
\sqrt{n} \,\, SR   \,\, \underbrace{\sqrt{ 1 + \frac{ 2 \rho }{ 1-\rho} \left( 1 - \frac{ 1 - \rho^{q} }{ q  (1-\rho)} \right) }}_{= \delta}
\end{equation}
follows a Student distribution with degree of freedom $n-1$ and non centrality parameter given by
\begin{equation}\label{eq:nonCentralityParam}
\sqrt{n} \,\, SR  \,\,  \delta
\end{equation}
\end{proposition}

\begin{proof}
See appendix section \ref{proof1}
\end{proof}

\begin{remark}
Proposition \ref{prop1} provides a nice and intuitive test for the significance of the Sharpe ratio. However, one need for the Student correction to compute the  first order auto-correlation  $\rho$ which can be tricky on data. A good recipe to infer auto correlation from data is to average the computation over the first order auto-correlation inferred from the lag one, two and three time series as follows. First compute for the time serie $(R_t)_{t=1, \ldots, n}$ the first order auto-correlation with the lag one time serie $R_{t-1} $:
\begin{equation}\label{eq:autocorrel1}
\rho_{1} = \frac{ \Cov( R_t, R_{t-1} ) } { \left(  \Var(R_t) + \Var(R_{t-1})  \right) / 2 }
\end{equation}

Compute the first order auto-correlation with the lag two time serie $R_{t-2} $:

\begin{equation}\label{eq:autocorrel2}
\rho_{2} = \left( \frac{ | \Cov( R_t, R_{t-2} ) |_{+} } { \left(  \Var(R_t) + \Var(R_{t-2})  \right) / 2 } \right)^{1 / 2} 
\end{equation}
where in equation \eqref{eq:autocorrel2}, we have force that the numerator is positive thanks to the positive value $ |x  |_{+} = \max( 0, x)$ in order for the square root to be real. 
We can also compute the first order auto-correlation with the lag three time serie $R_{t-3} $ as follows:

\begin{equation}\label{eq:autocorrel3}
\rho_{3} = \left( \frac{ \Cov( R_t, R_{t-2} ) } { \left(  \Var(R_t) + \Var(R_{t-2})  \right) / 2 } \right)^{1 / 3} 
\end{equation}

In all these expressions, the denominator term representing the variance of the time series is symmetrized to make formulae more robust.
Finally, we compute the \textit{true} first order auto-correlation as the average of $\rho_{1}, \rho_{2}, \rho_{3}$
$$
\rho = \frac{ \rho_{1} + \rho_{2} + \rho_{3}}{3}
$$
\end{remark}

\begin{proposition}\label{prop2} \textbf{Fisher and Beta Distributions} - 
A straight consequence of proposition \ref{prop1} is that the 
\begin{itemize}
\item the \textit{Studentized} F-statistic
$$
n \,\, SR^2 \,\,  \delta^2
$$ follows a Fisher–Snedecor distribution with parameters $d_1=1$ and $d_2=n-1$. This allows to convert the Student test to a Fisher test
\item the \textit{Studentized} Beta-statistic 
$$
\frac{ 1}{ 1 + SR^2 \,\,  \delta^2}
$$ follows a Beta distribution with parameters $d_1=(n-1) / 2$ and $d_2=1 / 2$. This allows to convert the Student test to a Beta test
\end{itemize}
\end{proposition}

\begin{proof}
See appendix section \ref{proof2}
\end{proof}
\vspace{0.5cm}

\begin{proposition}\label{prop3} \textbf{Wald Statistic} - 
The asymptotic distribution  of the Fisher distribution is the Chi Square distribution with one degree of freedom. If we make the test directly on the non \textit{Studentized} F-statistic $n \,\, SR^2$, it converges for $n$ large to the square of the standard normal distribution $\mathcal{N}(0,1)$. This allows to convert the Student (Fisher and Beta) test to the Wald test. These two tests are asymptotically equivalent.
\end{proposition}

\begin{proof}
See appendix section \ref{proof3}
\end{proof}

\begin{proposition}\label{prop3_2} \textbf{Wald Statistic 2} - 
If we apply the same reasoning but to the \textit{Studentized} F-statistic $n \,\, SR^2 \,\,  \delta^2$ converges for $n$ large to the square of the standard normal distribution $\mathcal{N}(0,1)$. This allows to convert the Student (Fisher and Beta) test to the Wald test. These two tests are asymptotically equivalent.
\end{proposition}

\begin{proposition}\label{prop3bis} \textbf{Modified Asymptotic Distribution} - 
A better approximation for the asymptotic distribution of the Fisher one is to compute the following statistic
\begin{equation}\label{eq:Walck}
 \sqrt{n} \,\, \frac{ SR \,\, \delta  (1- \frac 1 {4(n-1)})  }{ \sqrt{1 + \frac{SR^2 \delta^2}{2 (1-1/n) }} } \rightarrow N(0, 1)
\end{equation}
\end{proposition}

\begin{proof}
See appendix section \ref{proof3bis}
\end{proof}

\begin{remark}
It is worth noticing that $SR  = O( \frac{1}{\sqrt n })$ as $n \to \infty$, which explains that the result is asymptotically equivalent to proposition \ref{prop3_2} .

\end{remark}

\section{Statistical Tests}
In this section, leveraging distributions provided in section \ref{sec:distributions}, we provide the corresponding statistical tests. These tests provide a statistical answer whether a manager has some real skill or just is lucky. For all these tests, we have the same null hypothesis.

\begin{definition} \textbf{General null hypothesis} - 
The test null hypothesis $H_0$ is that the obtained empirical Sharpe ratio is of zero mean. In mathematical terms, it says $\mathbb{E}[SR] = 0$ 
If we fail to reject the null hypothesis, we conclude that the Sharpe ratio is not statically significant. Conversely, if we reject the null hypothesis, we conclude that the Sharpe ratio is statically significant. At this stage, we have two ways to do the test. We can do a one-tailed or a two-tailed test.
\end{definition}

\begin{definition} \textbf{One-tailed test} - 
In one-tailed test, the null hypothesis is:
$$
H_0:  \mathbb{E}[SR] = 0 \qquad \mathrm{versus} \quad H_1 : \mathbb{E}[SR] > 0,
$$
For a significance level $\alpha$ (typically $\alpha=5 \%$), 
we reject $H_0$ if the statistic $\hat{SR}$ is greater than
the $1-\alpha$ quantile of the distribution of the Sharpe ratio. For a given observed value $s$ of our Sharpe ratio, the p-value is defined as $\mathbb{P}\left( SR > s \, |H_0 \right)$.
\end{definition}

\begin{definition} \textbf{Two-tailed test} - 
In two-tailed test, the null hypothesis is:
$$
H_0:  \mathbb{E}[SR] = 0 \qquad \mathrm{versus} \quad H_1 : \mathbb{E}[SR] \neq 0,
$$
Two-tailed test requires also that under null hypothesis, the distribution of the Sharpe ratio is symmetric. 
For a significance level $\alpha$ (typically $\alpha=5 \%$), 
we reject $H_0$ if the statistic $| \hat{SR} |$ is greater than
the $1-\alpha/2$ quantile of the distribution of the Sharpe ratio. For a given observed value $s >0$ of our Sharpe ratio, the p-value is defined as $\mathbb{P}\left( |SR |> s \, |H_0 \right) = 2 \mathbb{P}\left( SR > s \, |H_0 \right)$.
\end{definition}

\begin{definition} \textbf{Skill and Luck percentage} - 
Given a statistic $SR$ and its observed value $s$, we define the luck percentage $\Prob(L)$ as the p-value under the null hypothesis and the skill percentage $\Prob(S)$ as the complementary probability  $\Prob(S) = 1 - \Prob(L) $. 
For the one-tailed test, this boils down to
$$
 \Prob(L) = \Prob(SR > s \, |H_0 ) = 1-\Prob(SR < s \, |H_0 ), \qquad   \Prob(S) =\Prob(SR <  s\,  |H_0 )
$$
while for two-tailed test requiring a symmetric statistic, and $s >0$ this boils down to
$$
\Prob(L) = \Prob( |SR| > s \, |H_0 )  = 2 \left(1-\Prob(SR <s \,  |H_0 )\right),  \qquad  \Prob(S) =  2 \Prob( SR < s \, |H_0) - 1
$$
\end{definition}

\begin{remark}
The difference between one and two-tailed test is subtle. In two-tailed test (which is the most common statistical test), we do not make any hypothesis on the sign of the Sharpe ratio. Hence the p-value under the null hypothesis adds on purpose both the right and left cumulative part of the distribution below and above the observed realization of the statistic. To reject the null hypothesis, we compare the p-value with a confidence level $\alpha$. If the p-value is above the confidence level $\alpha$, we reject the null hypothesis. As the p-value in two-tailed test is greater than in one-tailed test (for the same distribution assumptions), two-tailed test leads to reject the null hypothesis more often. Conversely, we fail to accept the null hypothesis less often in two-tailed tests than one-tailed ones. This can be seen as a paradox at first sight, as the two-tailed test implicitly ensures that when we reject the null hypothesis, we can only conclude that the Sharpe ratio has a non null mean (the opposite of the null hypothesis). This does not confirm that the mean is positive. 
At the opposite, for the one-tailed test when we reject the null hypothesis, we explicitly conclude that the mean of the Sharpe ratio is positive. But we have to pay attention that this type of test requires the mean of the Sharpe to be non negative. So we reject more often (than the two-tailed test) the hypothesis that the Sharpe ratio has a zero mean but, paradoxically, we already know that the mean cannot be negative, which is a huge bias in our analysis of the Sharpe ratio. 
\end{remark}

\begin{remark}
There are multiple choices for the distribution of the Sharpe. If we take the real distribution of the Sharpe, under certains assumptions, we can prove that it is a Student distribution. This distribution leads to a Student test that can be one or two-tailed. If we look at the Fisher statistic given by $N \,\, \delta^2 \,\, \frac{s^2}{F}$ it is necessarily a two-tailed test. The same applies for the Wald test as it takes the asymptotic distribution of the Fisher one, that is a chi-squared distribution. For implementation purposes, we can take the square root of the Wald statistic that is then a Gaussian but with the restriction that a one-tailed test does not make sense. Hence like for Fisher, only two-tailed tests are permitted.
\end{remark}

\begin{remark}
Last but not least, if we wanted to have a null hypothesis of the form 
$$
H_0: \mathbb{E}[SR] > \mu_0 \quad \mathrm{with} \quad \mu_0>0
$$
we would need to add an explicit assumption on the distribution of the first moment of the Sharpe ratio. This would require to make a \textit{prior} assumption on the parameters of the Sharpe ratio's distribution, leading us to do Bayesian statistics. We leave this for further study as one needs some good assumptions on the prior, which is not an obvious choice.
\end{remark}

\begin{proposition}\label{prop4} \textbf{Student Test} -
In the Student test, for $N=N_{d}$ daily trades (resp. $N=N^{m}$ monthly trades) corresponding to an annual factor $F=252$ (resp. $F=12$), the p-value is computed for one-tailed test as:
$$
\Prob(L) = \Prob(SR >s \, | H_0) = 1-\mathcal{T}^{cdf}\left( \sqrt{N} \,\, \delta \,\, \frac{s}{\sqrt{F}}\right)  
$$
for two-tailed test as:
$$
\Prob(L) = \Prob(|SR| >s \, | H_0) = 2- 2\, \mathcal{T}^{cdf}\left( \sqrt{N} \,\, \delta \,\, \frac{s}{\sqrt{F}}\right)  
$$
where $\mathcal{T}^{cdf}$ is the Student cumulated density function with $N-1$ degree of freedom. 
For a level of confidence $C$, the corresponding annualized Sharpe is given for daily calculations (resp. monthly calculations) for one-tailed test by:
\begin{equation}\label{eq:StudentEq}
\frac{ \sqrt{252}}{ \delta \sqrt{N^{d}} } \mathcal{T}^{inv}(N^{d}-1,  C)\qquad   \qquad \mathrm{resp.} \quad  \frac{ \sqrt{12}}{ \delta \sqrt{N^{m}} } \mathcal{T}^{inv}(N^{m}-1, C)
\end{equation}
and for two-tailed test by:
\begin{equation}\label{eq:StudentEq2}
\frac{ \sqrt{252}}{ \delta \sqrt{N^{d}} } \mathcal{T}^{inv}(N^{d}-1, \frac{1+ C}{2})\qquad   \qquad \mathrm{resp.} \quad  \frac{ \sqrt{12}}{ \delta \sqrt{N^{m}} } \mathcal{T}^{inv}(N^{m}-1, \frac{1+ C}{2})
\end{equation}
where $ \mathcal{T}^{inv}(n-1, x)  $ is the inverse of the cumulated density function of a student distribution with degree of freedom $n-1$ for a cumulated probability $0 \leq x \leq 1$.
\end{proposition}

\begin{proof}
Immediate using previous results.
\end{proof}

\begin{proposition}\label{prop5} \textbf{Fisher Tests} - 
In the Fisher test, the p-value is computed as follows:
$$
\Prob(L) =  \Prob(SR >s \, | H_0) =  1 - \mathcal{F}^{cdf}\left( N \,\, \delta^2 \,\, \frac{s^2}{F}\right)  
$$
where $\mathcal{F}^{cdf}$ is the Fisher cumulated density function with $1, N-1$ parameters.  
\end{proposition}

\begin{proof}
Immediate using previous results.
\end{proof}

\begin{proposition}\label{prop6} \textbf{Wald Test} -
In the Wald test, the p-value is computed with a two-tailed test and is given by:
$$
\Prob(L) =  1 - \Prob(|SR| < s\, | H_0) =  2-2 \, \mathcal{N}^{cdf}\left( s \,\, \frac{\sqrt{N}}{ \sqrt{F} }\right)   
$$
where $\mathcal{N}^{cdf}$ is the normal  cumulated density function for the standard normal distribution $\mathcal{N}(0,1)$. 
Hence for a level of confidence $C$, the corresponding annualized Sharpe for daily calculations (resp. monthly calculations) is given by
\begin{equation}\label{eq:WaldEq}
\frac{ \sqrt{252}}{\sqrt{N^{d}} } \mathcal{N}^{inv}\left(\frac{1 + C}{2}\right)\qquad   \qquad \mathrm{resp.} \quad  \frac{ \sqrt{12}}{\sqrt{N^{m}} } \mathcal{N}^{inv}\left(\frac{1 + C}{2}\right)
\end{equation}
where $ \mathcal{N}^{inv}(x)$ is the inverse of the cumulated density function of a standard normal $\mathcal{N}(0,1)$ for a cumulated probability $0 \leq x \leq 1$ : 
$\int_{\infty}^{\mathcal{N}^{inv}(x)} N(0,1) du = x$. 
\end{proposition}

\begin{proof}
See appendix section \ref{proof6}
\end{proof}

\begin{proposition}\label{prop6bis} \textbf{Other Wald Test} -
Another form of the Wald test is to apply the \textit{Studentized}  version of our Sharpe ratio as follows:
$$
\Prob(L) = 1 - \Prob(|SR| < s\, | H_0)  =  2-2 \, \mathcal{N}^{cdf}\left( s \,\, \frac{\sqrt{N} \delta}{ \sqrt{F}}  \right)    
$$
where $\mathcal{N}^{cdf}$ is the normal  cumulated density function for the standard normal distribution. 
Hence for a level of confidence $C$, the corresponding annualized Sharpe is given by
\begin{equation}\label{eq:WaldEq2}
\frac{ \sqrt{252}}{\delta \sqrt{N^{d}} } \mathcal{N}^{inv}\left(\frac{1 + C}{2}\right)\qquad   \qquad \mathrm{resp.} \quad  \frac{ \sqrt{12}}{\delta \sqrt{N^{m}} } \mathcal{N}^{inv}\left(\frac{1 + C}{2}\right)
\end{equation}
\end{proposition}

\begin{proposition}\textbf{Other Wald Test 2 } -
Another form of the Wald test is to use proposition \ref{prop3bis} and correct the statistics so that
\begin{equation}
\Prob(L) =  1 - \Prob(|SR| < s\, | H_0) = 2 - 2  \, \mathcal{N}^{cdf}\left(  s \,\,   \frac{\sqrt{N} \delta}{  \sqrt{F}}  \,\, \frac{1- \frac 1 {4(N-1)}}{ \sqrt{1 + \frac{s^2 \delta^2}{2 (1-1/N) }} }  \right)   
\end{equation}
where $\mathcal{N}^{cdf}$ is the normal  cumulated density function for the standard normal distribution. 
Hence for a level of confidence $L$, the corresponding annualized Sharpe is given by
\begin{equation}\label{eq:WaldEq3}
\frac{ \sqrt{F} }{\delta \sqrt{N} } \,\, \frac{1}{\sqrt{ \left( 1- \frac{1}{4(N-1)} \right)^2 - \frac{F}{N} \frac{ \left( \mathcal{N}^{inv}\left(\frac{1 + C}{2}\right) \right)^2 }{2 (1-1/N)}} }  \,\, \mathcal{N}^{inv}\left(\frac{1 + C}{2}\right)
\end{equation}
\end{proposition}

\section{Statistical Tables}
In this section, we provide various statistical tables to test the significance of a Sharpe ratio.
\subsection{Color codes}
When providing a sharpe ratio, we have used color codes to identify quickly sharpe ratio values and probabilities. The various thresholds for Sharpe ratios, summarized by table \ref{tab:tabSRColor}, are not chosen randomly. They corresponds to $0.50, 1.00, 1.50$ and $2.00$ as these levels of Sharpe ratios are within the range of low to medium frequency funds. For the skill probabilities, summarized in  table \ref{tab:tabProbaColor}, we have used the following thresholds: $80 \%, 90\%, 95 \%, 97.5 \%$ and $99\%$ as they are also meaningful from a statistical point of view. $90\%, 95 \%$ and $97.5 \%$ are typically what is considered to be "closed" to certitude from a statistical point of view taking into natural noise in data. 

\begin{multicols}{2}

\begin{table}[H]
  \centering
  \caption{Color code for Sharpe ratio}
    \begin{tabular}{|c|}
    \midrule
    \rowcolor[rgb]{ 1,  .247,  .247}   $SR >  2.00$  \\
    \midrule
    \rowcolor[rgb]{ 1,  .475,  .475}   $2.00 \geq SR > 1.50$  \\
    \midrule
    \rowcolor[rgb]{ 1,  .686,  .686}   $1.50 \geq SR > 1.00$   \\
    \midrule
    \rowcolor[rgb]{ 1,  .922,  .922}   $1.00 \geq SR > 0.50$   \\
    \midrule
         $0.50 \geq SR$ \\
    \midrule
    \end{tabular}%
  \label{tab:tabSRColor}%
\end{table}%

\vfill\null

\columnbreak

\begin{table}[H]
  \centering
  \caption{Color code for skill probability}
    \begin{tabular}{|c|}
    \midrule
                                              $00\% \leq \mathbb{P}(.) < 80\%$ \\
    \midrule
    \rowcolor[rgb]{ .788,  1,  .882} $80\% \leq \mathbb{P}(.) < 90\%$ \\
    \midrule
    \rowcolor[rgb]{ .592,  1,  .776} $90\% \leq \mathbb{P}(.) < 95\%$ \\
    \midrule
    \rowcolor[rgb]{ .114,  1,  .514} $95\% \leq \mathbb{P}(.) < 97.5\%$ \\
    \midrule
    \rowcolor[rgb]{ 0,  .855,  .388} $97.5\% \leq \mathbb{P}(.) < 99\%$ \\
    \midrule
    \rowcolor[rgb]{ 0,  .69,  .314} $99\% \leq \mathbb{P}(.) $ \\
    \midrule
    \end{tabular}%
  \label{tab:tabProbaColor}%
\end{table}%
\end{multicols}

\subsection{Sharpe for 90\% skill target}
\subsubsection{Wald assumptions}
Tables \ref{tab:tab1} and \ref{tab:tab2} provide the minimum Sharpe level for a 90\% skill percentage, under various auto-correlation assumptions. Table \ref{tab:tab1} shows the results for daily periods while table \ref{tab:tab2} gives the same information but for monthly periods. Tables \ref{tab:tab1} and \ref{tab:tab2} compute the skill percentage using the Wald test, hence assuming that 
$$
SR \,\, \frac{\sqrt{N} \delta}{ \sqrt{F}} \sim \mathcal{N}(0,1)
$$
with a two-tailed test method. Let us explain what table \ref{tab:tab1}'s numbers mean by taking various examples. To ensure a $90\%$ skill percentage, we need for 150 daily periods a Sharpe ratio at least equal to $2.9$  for a correlation of $-30\%$ and only to $1.57$ for a correlation of $+30\%$. If we double the daily period (300 daily periods), we need  a Sharpe ratio at least equal to $2.05$ for a correlation of $-30\%$ and only to $1.11$ for a correlation of $+30\%$. Notice that Sharpe decreases in square root of time $\sqrt{N}$. Hence to reduced the required Sharpe ratio by two, we need a daily period $4$ times longer. This is confirmed by the minimum Sharpe ratio, for a period of 600 daily periods, which is equal to $1.45$ for a correlation of $-30\%$ and only to $0.78$ for a correlation of $+30\%$. Last but not least, we see that the more auto correlation, the lower the required Sharpe ratio.

Table \ref{tab:tab2} provides similar results that are very comparable to daily computations. For instance, a two year period would corresponds to 24 months and roughly 500 daily observations (rigorously, $2 \times 252$). Hence for 24 months, to ensure a $90\%$ skill percentage, we need a Sharpe ratio at least equal to $1.54$  for a correlation of $-30\%$, $1.16$ for a zero correlation and to $0.88$ for a correlation of $+30\%$. This is comparable to a minimum Sharpe ratio at least equal to $1.59$  for a correlation of $-30\%$, $1.17$ for a zero correlation and to $0.86$ for a correlation of $+30\%$ for $500$ daily periods. There is however a slight difference between monthly and daily periods. To account for a daily period, we need on a given day at least one trade. Likewise, to account for a month, we need on a given month at least one trade. When looking at intra-day strategies, there is no guarantee that there are trades on a given day if we do not use a large number of different strategies while there is little chance that we skip a month.

\subsubsection{Student assumptions}
Tables \ref{tab:tab3} and \ref{tab:tab4} provide the minimum Sharpe level for a 90\% skill percentage, under various auto-correlation assumptions. Table \ref{tab:tab3} shows the results for daily periods while table \ref{tab:tab4} gives the same information but for monthly periods. Tables  \ref{tab:tab3} and \ref{tab:tab4} compute the skill percentage using the Student test, hence assuming that 
$$
SR \,\, \frac{\sqrt{N} \delta}{ \sqrt{F}} \sim \mathcal{T}(N-1)
$$
with a one-tailed test method. \\
The skill percentages for two-tailed method are expressed in tables \ref{tab:tab3bis} and \ref{tab:tab4bis} for daily and monthly periods.
It is worth mentioning that there are large differences between one and two-tailed tests. For instance, to ensure a $90\%$ skill percentage for a one-tailed test (table \ref{tab:tab3}) and 500 daily periods, we need a Sharpe ratio at least equal to $1.24$  for a correlation of $-30\%$, $0.91$ for a zero correlation and to $0.67$ for a correlation of $+30\%$, which are comparable to the numbers for the same correlation levels but for $800$ daily periods ($1.26$ for a correlation of $-30\%$, $0.92$ for a zero correlation and to $0.68$ for a correlation of $+30\%$, see table \ref{tab:tab3bis}). From a practical point of view, we need almost twice as much time to ensure the same skill level between one and two-tailed tests!

\subsection{95\% skill target}
\subsubsection{Wald assumptions}
Tables \ref{tab:tab5} and \ref{tab:tab6} provide the minimum Sharpe level for a 95\% skill percentage, under various auto-correlation assumptions. Table \ref{tab:tab5} shows the results for daily periods while table \ref{tab:tab6} gives the same information but for monthly periods. Tables  \ref{tab:tab5} and \ref{tab:tab6} compute the skill percentage using the Wald test, hence assuming that 
$$
SR \,\, \frac{\sqrt{N} \delta}{ \sqrt{F}} \sim \mathcal{N}(0,1)
$$
with a two-tailed test method.
\subsubsection{Student assumptions}
Tables \ref{tab:tab7} and \ref{tab:tab8} provide the minimum Sharpe level for a 95\% skill percentage, under various auto-correlation assumptions. Table \ref{tab:tab7} shows the results for daily periods while table \ref{tab:tab8} gives the same information but for monthly periods. Tables \ref{tab:tab7} and \ref{tab:tab8} compute the skill percentage using the Student test, hence assuming that 
$$
SR \,\, \frac{\sqrt{N} \delta}{ \sqrt{F}} \sim \mathcal{T}(N-1)
$$
with a one-tailed test method. \\
The skill percentages for two-tailed method are expressed in tables \ref{tab:tab7bis} and \ref{tab:tab8bis} for daily and monthly periods.
Again to make sure these tables are easy to understand, let us take some examples. Table \ref{tab:tab3} says that we need a Sharpe ratio at least equal to $1.50$  for a correlation of $-30\%$, $1.13$ for a zero correlation and to $0.85$ for a correlation of $+30\%$ for a period of three years (36 months). Same comments apply as before. Required Sharpe ratio decreases in square root of time.  And the larger the auto correlation, the lower the required Sharpe ratio.

\subsection{Sharpe ratio significance level}
In this part, we provide the minimum Sharpe ratio level to achieve a confidence level measured by the percentage of skill assuming zero correlation.
Tables \ref{tab:tab9} and \ref{tab:tab10} assume Wald test distribution (the normality of the Sharpe ratio). 
Table \ref{tab:tab9} looks at daily periods while table \ref{tab:tab10} at monthly periods for the one-tailed test.
Tables \ref{tab:tab11} and \ref{tab:tab12} assume Student one-tailed test distribution. 
Table \ref{tab:tab11} looks at daily periods while table \ref{tab:tab12} at monthly periods for the one-tailed test.
Tables \ref{tab:tab11bis} and \ref{tab:tab12bis} assume Student two-tailed test distribution. 
Table \ref{tab:tab11bis} looks at daily periods while table \ref{tab:tab12bis} at monthly periods for the one-tailed test.

\begin{remark}
Tables \ref{tab:tab3bis} and \ref{tab:tab7} are rigorously the same. A two-tailed Student test is equal to the one-tailed Student test whenever equations \eqref{eq:StudentEq} and \eqref{eq:StudentEq2} are equal. Denoting by $C_1=90 \%$, and $C_2=95 \%$, the two-tailed test computes 
$$
\frac{ \sqrt{252}}{ \delta \sqrt{N^{d}} } \mathcal{T}^{inv}(N^{d}-1, \frac{1+ C_1}{2})
$$
while the one-tailed test uses
$$
\frac{ \sqrt{252}}{ \delta \sqrt{N^{d}} } \mathcal{T}^{inv}(N^{d}-1, C_2)
$$
This explains why the two daily tables are the same. The same applies for table \ref{tab:tab4bis} and \ref{tab:tab8} .
\end{remark}

\subsection{Percentage of skill for a given level of Sharpe ratio}

In this section, we provide the percentage of skill for a given level of Sharpe ratio using the formulae under the Wald assumptions (two-tailed test) 
$$
\Prob(S) = \Prob(T < | t |) = 2 \mathcal{N}^{cdf}\left( \sqrt{N} \,\, \delta \,\, \frac{SR}{\sqrt{F}}\right) - 1
$$
and under the Student assumptions with one-tailed test:
$$
\Prob(S) = \Prob(T < t) = \mathcal{T}^{cdf}\left( \sqrt{N} \,\, \delta \,\, \frac{SR}{\sqrt{F}}\right)  
$$
and under the Student assumptions with two-tailed test:
$$
\Prob(S) = \Prob(T < | t |) = 2 \mathcal{T}^{cdf}\left( \sqrt{N} \,\, \delta \,\, \frac{SR}{\sqrt{F}}\right) - 1
$$
\subsubsection{Sharpe ratio of 0.5}
Tables \ref{tab:tab13} and \ref{tab:tab14} assume Wald test distribution (the normality of the Sharpe ratio). 
Table \ref{tab:tab13} looks at daily periods while table \ref{tab:tab14} at monthly periods for the one-tailed test.
Tables \ref{tab:tab15} and \ref{tab:tab16} assume Student one-tailed test distribution. 
Table \ref{tab:tab15} looks at daily periods while table \ref{tab:tab16} at monthly periods for the one-tailed test.
Tables \ref{tab:tab15bis} and \ref{tab:tab16bis} assume Student two-tailed test distribution. 
Table \ref{tab:tab15bis} looks at daily periods while table \ref{tab:tab16bis} at monthly periods for the one-tailed test.

\subsubsection{Sharpe ratio of 1}
Tables \ref{tab:tab17} and \ref{tab:tab18} assume Wald test distribution (the normality of the Sharpe ratio). 
Table \ref{tab:tab17} looks at daily periods while table \ref{tab:tab18} at monthly periods for the one-tailed test.
Tables \ref{tab:tab19} and \ref{tab:tab20} assume Student one-tailed test distribution. 
Table \ref{tab:tab19} looks at daily periods while table \ref{tab:tab20} at monthly periods for the one-tailed test.
Tables \ref{tab:tab19bis} and \ref{tab:tab20bis} assume Student two-tailed test distribution. 
Table \ref{tab:tab19bis} look at daily periods while table \ref{tab:tab20bis} at monthly periods for the one-tailed test.

\subsubsection{Sharpe ratio of 1.5}
Tables \ref{tab:tab21} and \ref{tab:tab22} assume Wald test distribution (the normality of the Sharpe ratio). 
Table \ref{tab:tab21} look at daily periods while table \ref{tab:tab22} at monthly periods for the one-tailed test.
Tables \ref{tab:tab23} and \ref{tab:tab24} assume Student one-tailed test distribution. 
Table \ref{tab:tab23} look at daily periods while table \ref{tab:tab24} at monthly periods for the one-tailed test.
Tables \ref{tab:tab23bis} and \ref{tab:tab24bis} assume Student two-tailed test distribution. 
Table \ref{tab:tab23bis} look at daily periods while table \ref{tab:tab24bis} at monthly periods for the one-tailed test.

\subsubsection{Sharpe ratio of 2}
Tables \ref{tab:tab25} and \ref{tab:tab26} assume Wald test distribution (the normality of the Sharpe ratio). 
Table \ref{tab:tab25} looks at daily periods while table \ref{tab:tab26} at monthly periods for the one-tailed test.
Tables \ref{tab:tab27} and \ref{tab:tab28} assume Student one-tailed test distribution. 
Table \ref{tab:tab27} looks at daily periods while table \ref{tab:tab28} at monthly periods for the one-tailed test.
Tables \ref{tab:tab27bis} and \ref{tab:tab28bis} assume Student two-tailed test distribution. 
Table \ref{tab:tab27bis} looks at daily periods while table \ref{tab:tab28bis} at monthly periods for the one-tailed test.

\section{Conclusion}
In this paper, we have provided the framework for validating the significance of a historically realized Sharpe with respect to statistical test. This allows us to define the percentage of skill and luck of a manager. This allows answering the question whether a manager was indeed lucky and its generated Sharpe is statistically significant or not. We provide statistical tables that provides guidelines in terms of statistical significance of a Sharpe ratio given correlation and number of observation.

\bibliographystyle{apalike}
\bibliography{mybib}

\clearpage

\section{Appendix}

\subsection{Proof of proposition \ref{prop:1}} \label{proof:1}
The denominator of the Sharpe can be computed as follows:

\begin{eqnarray}
Var[R_t(q)]  &= & \sum_{i=0}^{q-1} \sum_{j=0}^{q-1} Cov( R_{t-i}, R_{t-j} ) \\
&= & \sum_{i=0}^{q-1} Var(R_{t-i} ) + 2  \sum_{i=0}^{q-1} \sum_{j=i+1}^{q-1} Cov( R_{t-i}, R_{t-j} )
\end{eqnarray}

Using the change of variable, $k=j-i$, denoting by $\rho_{u,v} = Corr(R_{u}, R_{v})$ and by $\sigma^2_{u}=\Var(Var(R_{u})$, we get the following expression:
\begin{equation} \label{GeneralEqVar}
Var[R_t(q)]  =  \sum_{i=0}^{q-1} \sigma^2_{t-i} + 2  \sum_{k=1}^{q-1} \sum_{i = 0 }^{q-1-k } \rho_{t-i, t-i-k} \sigma_{t-i} \sigma_{t-i-k}
\end{equation}

Computed the fraction $\frac{SR(q)}{SR}$ leads to the result of equation \ref{SRq_eq1}. This is the most general formula that extends the one of  \cite{Lo_2002}. 

If the return process is stationary with a constant variance, then
\begin{itemize}
\item $\sigma_{u}$ does not depend on $u$ and can be denoted by $\sigma$, 
\item $\rho_{u,v}$ only depends on the absolute difference between $u$ and $v$ and is written for $v \geq u$ as $\rho_{v-u}$
\end{itemize}

Then equation \ref{GeneralEqVar} becomes
\begin{eqnarray}\label{GeneralEqVar2}
Var[R_t(q)]  &= &  \sigma^2  \left( q + 2  \sum_{k=1}^{q-1} (q-k) \rho_{k} \right)
\end{eqnarray}

Again, computing the fraction $\frac{SR(q)}{SR}$ leads to the result of equation \ref{SRq_eq2}. 

For AR(1) process, we have $\rho_k=\rho^k$ and  equation \ref{GeneralEqVar2} becomes
\begin{eqnarray}
Var[R_t(q)]  &= &  \sigma^2  \left[ q + \frac{ 2 \rho }{ 1-\rho} \left( 1 - \frac{ 1 - \rho^q}{q (1-\rho)} \right) \right]
\end{eqnarray}

where we have used $\sum_{k=0}^{q-2}  \rho^k = \frac{1- \rho^{q-1} }{1-\rho}$ 
and $\sum_{k=0}^{q-1} k \rho^{k-1} = \frac{ \frac{1- \rho^{q}}{1-\rho} - q \rho^{q-1}}{ (1-\rho) }$. 
Equation \ref{SRq_eq4} is immediate.

\subsection{Proof of proposition \ref{prop1}} \label{proof1}
The result is trivial for i.i.d. returns. For the case of AR(1) assumptions, we can apply previously results and correct by the auto correlation effect

\subsection{Proof of proposition \ref{prop2}} \label{proof2}
The result is trivial as if 
\begin{itemize}
\item If $X$ has a Student's t-distribution with degree of freedom $\nu$ then $Y=X^2$  has an F-distribution: $Y \sim \mathcal{F}\left(\nu_1 = 1, \nu_2 = \nu\right)$
\item If $X$ has a Student's t-distribution with degree of freedom $\nu$ then $Z=\frac{\nu}{\nu + X^2}$ has a beta-distribution  $Z \sim \mathcal{B}\left(\frac{\nu}2, \frac12\right)$
\end{itemize}

\subsection{Proof of proposition \ref{prop3}} \label{proof3}
Using the fact that the Fisher distribution $\lim_{n\to\infty} \mathcal{F}(1,n) =  \mathcal{N}(0,1)$, 
we have that $n \,\, SR^2 \,\,  \delta^2$ converges to a normal distribution. 
Hence, the Student and Fisher tests are asymptotically equivalent to the Wald test.

\subsection{Proof of proposition \ref{prop3bis}} \label{proof3bis}
When $n$ tends to infinity, the t-distribution denoted by $t$ tends to a normal distribution denoted by $N( 0, \sigma) $ whose variance is given by the variance of the Student distribution. Using the result of \cite{Walk_2007}, we conclude with
\begin{equation}\label{Walk_result}
\frac{t (1- \frac 1 {4(n-1)}) }{ \sqrt{1 + \frac{t^2}{2 (n-1) }} } \rightarrow N(0, 1)
\end{equation}

\subsection{Proof of proposition \ref{prop6}} \label{proof6}
Immediate using previous results noticing that this is a two-tailed test and that inverting the relationship 
$$
\Prob(S) =  \Prob(T < |t|) = 2 \Prob(T < t) - 1
$$
implies 
$$
t = CDF^{-1}\left( \frac{1+\Prob(S) }{2} \right)
$$

\pagebreak
\section{Statistical Tables}\label{statistical_tables}
\subsection{Sharpe for 90\% skill target}
\subsubsection{Wald assumptions}
\vspace{-0.5cm}

\begin{table}[H]
  \centering
  \caption{Sharpe level for 90\% targeted skill using Wald two-tailed test for \textbf{daily} periods under various auto-correlation assumptions}
  \vspace{0.25cm}
  \resizebox {\textwidth} {!} {
\begin{tabular}{|r|r|r|r|r|r|r|r|r|r|r|r|r|r|r|r|r|r|r|r|r|r|}
    \toprule
    \multicolumn{1}{|p{5.5em}|}{\textbf{ \textbackslash{} Observ.\newline{}Correl.}} & \textbf{25} & \textbf{50} & \textbf{100} & \textbf{150} & \textbf{200} & \textbf{250} & \textbf{300} & \textbf{350} & \textbf{400} & \textbf{450} & \textbf{500} & \textbf{550} & \textbf{600} & \textbf{650} & \textbf{700} & \textbf{750} & \textbf{800} & \textbf{850} & \textbf{900} & \textbf{950} & \textbf{1000} \\
    \midrule
    \textbf{-90\%} & \cellcolor[rgb]{ 1,  .247,  .247}       22.35  & \cellcolor[rgb]{ 1,  .247,  .247}      15.80  & \cellcolor[rgb]{ 1,  .247,  .247}    11.17  & \cellcolor[rgb]{ 1,  .247,  .247}       9.12  & \cellcolor[rgb]{ 1,  .247,  .247}       7.90  & \cellcolor[rgb]{ 1,  .247,  .247}       7.07  & \cellcolor[rgb]{ 1,  .247,  .247}       6.45  & \cellcolor[rgb]{ 1,  .247,  .247}       5.97  & \cellcolor[rgb]{ 1,  .247,  .247}       5.59  & \cellcolor[rgb]{ 1,  .247,  .247}       5.27  & \cellcolor[rgb]{ 1,  .247,  .247}       5.00  & \cellcolor[rgb]{ 1,  .247,  .247}       4.76  & \cellcolor[rgb]{ 1,  .247,  .247}       4.56  & \cellcolor[rgb]{ 1,  .247,  .247}       4.38  & \cellcolor[rgb]{ 1,  .247,  .247}       4.22  & \cellcolor[rgb]{ 1,  .247,  .247}       4.08  & \cellcolor[rgb]{ 1,  .247,  .247}       3.95  & \cellcolor[rgb]{ 1,  .247,  .247}       3.83  & \cellcolor[rgb]{ 1,  .247,  .247}       3.72  & \cellcolor[rgb]{ 1,  .247,  .247}       3.63  & \cellcolor[rgb]{ 1,  .247,  .247}       3.53  \\
\end{tabular}%
}
  \label{tab:tab1}%
\end{table}%

\begin{table}[H]
  \centering
  \caption{Sharpe level for 90\% targeted skill using Wald two-tailed test for \textbf{monthly} periods under various auto-correlation assumptions}
  \vspace{0.25cm}
  \resizebox {\textwidth} {!} {
    \begin{tabular}{|r|r|r|r|r|r|r|r|r|r|r|r|r|r|r|r|r|r|r|r|r|}
    \toprule
    \multicolumn{1}{|p{5.5em}|}{\textbf{ \textbackslash{} Observ.\newline{}Correl.}} & \textbf{6} & \textbf{12} & \textbf{18} & \textbf{24} & \textbf{30} & \textbf{36} & \textbf{42} & \textbf{48} & \textbf{54} & \textbf{60} & \textbf{66} & \textbf{72} & \textbf{78} & \textbf{84} & \textbf{90} & \textbf{96} & \textbf{102} & \textbf{108} & \textbf{114} & \textbf{120} \\
    \midrule
    \textbf{-90\%} & \cellcolor[rgb]{ 1,  .247,  .247}         8.10  & \cellcolor[rgb]{ 1,  .247,  .247}         5.73  & \cellcolor[rgb]{ 1,  .247,  .247}       4.68  & \cellcolor[rgb]{ 1,  .247,  .247}       4.05  & \cellcolor[rgb]{ 1,  .247,  .247}       3.62  & \cellcolor[rgb]{ 1,  .247,  .247}       3.31  & \cellcolor[rgb]{ 1,  .247,  .247}       3.06  & \cellcolor[rgb]{ 1,  .247,  .247}       2.86  & \cellcolor[rgb]{ 1,  .247,  .247}       2.70  & \cellcolor[rgb]{ 1,  .247,  .247}       2.56  & \cellcolor[rgb]{ 1,  .247,  .247}       2.44  & \cellcolor[rgb]{ 1,  .247,  .247}       2.34  & \cellcolor[rgb]{ 1,  .247,  .247}       2.25  & \cellcolor[rgb]{ 1,  .247,  .247}       2.17  & \cellcolor[rgb]{ 1,  .247,  .247}       2.09  & \cellcolor[rgb]{ 1,  .247,  .247}       2.03  & \cellcolor[rgb]{ 1,  .475,  .475}       1.96  & \cellcolor[rgb]{ 1,  .475,  .475}       1.91  & \cellcolor[rgb]{ 1,  .475,  .475}       1.86  & \cellcolor[rgb]{ 1,  .475,  .475}       1.81  \\
\end{tabular}%
}
  \label{tab:tab2}%
\end{table}%

\pagebreak
\subsubsection{Student assumptions and one-tailed test}
\vspace{-0.5cm}

\begin{table}[H]
  \centering
  \caption{Sharpe level for 90\% targeted skill using Student one-tailed test for \textbf{daily} periods under various auto-correlation assumptions}
  \vspace{0.25cm}
  \resizebox {\textwidth} {!} {
    \begin{tabular}{|r|r|r|r|r|r|r|r|r|r|r|r|r|r|r|r|r|r|r|r|r|r|}
    \toprule
    \multicolumn{1}{|p{5.5em}|}{\textbf{ \textbackslash{} Observ.\newline{}Correl.}} & \textbf{25} & \textbf{50} & \textbf{100} & \textbf{150} & \textbf{200} & \textbf{250} & \textbf{300} & \textbf{350} & \textbf{400} & \textbf{450} & \textbf{500} & \textbf{550} & \textbf{600} & \textbf{650} & \textbf{700} & \textbf{750} & \textbf{800} & \textbf{850} & \textbf{900} & \textbf{950} & \textbf{1000} \\
    \midrule
    \textbf{-90\%} & \cellcolor[rgb]{ 1,  .247,  .247}       17.90  & \cellcolor[rgb]{ 1,  .247,  .247}      12.48  & \cellcolor[rgb]{ 1,  .247,  .247}       8.76  & \cellcolor[rgb]{ 1,  .247,  .247}       7.14  & \cellcolor[rgb]{ 1,  .247,  .247}       6.18  & \cellcolor[rgb]{ 1,  .247,  .247}       5.52  & \cellcolor[rgb]{ 1,  .247,  .247}       5.04  & \cellcolor[rgb]{ 1,  .247,  .247}       4.66  & \cellcolor[rgb]{ 1,  .247,  .247}       4.36  & \cellcolor[rgb]{ 1,  .247,  .247}       4.11  & \cellcolor[rgb]{ 1,  .247,  .247}       3.90  & \cellcolor[rgb]{ 1,  .247,  .247}       3.72  & \cellcolor[rgb]{ 1,  .247,  .247}       3.56  & \cellcolor[rgb]{ 1,  .247,  .247}       3.42  & \cellcolor[rgb]{ 1,  .247,  .247}       3.29  & \cellcolor[rgb]{ 1,  .247,  .247}       3.18  & \cellcolor[rgb]{ 1,  .247,  .247}       3.08  & \cellcolor[rgb]{ 1,  .247,  .247}       2.99  & \cellcolor[rgb]{ 1,  .247,  .247}       2.90  & \cellcolor[rgb]{ 1,  .247,  .247}       2.83  & \cellcolor[rgb]{ 1,  .247,  .247}       2.75  \\
\end{tabular}%
}
  \label{tab:tab3}%
\end{table}%

\pagebreak

\begin{table}[H]
  \centering
  \caption{Sharpe level for 90\% targeted skill using Student one-tailed test for \textbf{monthly} periods under various auto-correlation assumptions}
  \vspace{0.25cm}
  \resizebox {\textwidth} {!} {
    \begin{tabular}{|r|r|r|r|r|r|r|r|r|r|r|r|r|r|r|r|r|r|r|r|r|}
    \toprule
    \multicolumn{1}{|p{5.5em}|}{\textbf{ \textbackslash{} Observ.\newline{}Correl.}} & \textbf{6} & \textbf{12} & \textbf{18} & \textbf{24} & \textbf{30} & \textbf{36} & \textbf{42} & \textbf{48} & \textbf{54} & \textbf{60} & \textbf{66} & \textbf{72} & \textbf{78} & \textbf{84} & \textbf{90} & \textbf{96} & \textbf{102} & \textbf{108} & \textbf{114} & \textbf{120} \\
    \midrule
    \textbf{-90\%} & \cellcolor[rgb]{ 1,  .247,  .247}         7.27  & \cellcolor[rgb]{ 1,  .247,  .247}         4.75  & \cellcolor[rgb]{ 1,  .247,  .247}       3.79  & \cellcolor[rgb]{ 1,  .247,  .247}       3.25  & \cellcolor[rgb]{ 1,  .247,  .247}       2.89  & \cellcolor[rgb]{ 1,  .247,  .247}       2.63  & \cellcolor[rgb]{ 1,  .247,  .247}       2.42  & \cellcolor[rgb]{ 1,  .247,  .247}       2.26  & \cellcolor[rgb]{ 1,  .247,  .247}       2.13  & \cellcolor[rgb]{ 1,  .247,  .247}       2.02  & \cellcolor[rgb]{ 1,  .475,  .475}       1.92  & \cellcolor[rgb]{ 1,  .475,  .475}       1.84  & \cellcolor[rgb]{ 1,  .475,  .475}       1.77  & \cellcolor[rgb]{ 1,  .475,  .475}       1.70  & \cellcolor[rgb]{ 1,  .475,  .475}       1.64  & \cellcolor[rgb]{ 1,  .475,  .475}       1.59  & \cellcolor[rgb]{ 1,  .475,  .475}       1.54  & \cellcolor[rgb]{ 1,  .686,  .686}       1.50  & \cellcolor[rgb]{ 1,  .686,  .686}       1.46  & \cellcolor[rgb]{ 1,  .686,  .686}       1.42  \\
\end{tabular}%
}
  \label{tab:tab4}%
\end{table}%

\pagebreak

\subsubsection{Student assumptions and two-tailed test}
\vspace{-0.5cm}

\begin{table}[H]
  \centering
  \caption{Sharpe level for 90\% targeted skill using Student two-tailed test for \textbf{daily} periods under various auto-correlation assumptions}
  \vspace{0.25cm}
  \resizebox {\textwidth} {!} {
    \begin{tabular}{|r|r|r|r|r|r|r|r|r|r|r|r|r|r|r|r|r|r|r|r|r|r|}
    \toprule
    \multicolumn{1}{|p{5.5em}|}{\textbf{ \textbackslash{} Observ.\newline{}Correl.}} & \textbf{25} & \textbf{50} & \textbf{100} & \textbf{150} & \textbf{200} & \textbf{250} & \textbf{300} & \textbf{350} & \textbf{400} & \textbf{450} & \textbf{500} & \textbf{550} & \textbf{600} & \textbf{650} & \textbf{700} & \textbf{750} & \textbf{800} & \textbf{850} & \textbf{900} & \textbf{950} & \textbf{1000} \\
    \midrule
    \textbf{-90\%} & \cellcolor[rgb]{ 1,  .247,  .247}       23.24  & \cellcolor[rgb]{ 1,  .247,  .247}      16.11  & \cellcolor[rgb]{ 1,  .247,  .247}    11.28  & \cellcolor[rgb]{ 1,  .247,  .247}       9.18  & \cellcolor[rgb]{ 1,  .247,  .247}       7.94  & \cellcolor[rgb]{ 1,  .247,  .247}       7.09  & \cellcolor[rgb]{ 1,  .247,  .247}       6.47  & \cellcolor[rgb]{ 1,  .247,  .247}       5.99  & \cellcolor[rgb]{ 1,  .247,  .247}       5.60  & \cellcolor[rgb]{ 1,  .247,  .247}       5.28  & \cellcolor[rgb]{ 1,  .247,  .247}       5.01  & \cellcolor[rgb]{ 1,  .247,  .247}       4.77  & \cellcolor[rgb]{ 1,  .247,  .247}       4.57  & \cellcolor[rgb]{ 1,  .247,  .247}       4.39  & \cellcolor[rgb]{ 1,  .247,  .247}       4.23  & \cellcolor[rgb]{ 1,  .247,  .247}       4.09  & \cellcolor[rgb]{ 1,  .247,  .247}       3.96  & \cellcolor[rgb]{ 1,  .247,  .247}       3.84  & \cellcolor[rgb]{ 1,  .247,  .247}       3.73  & \cellcolor[rgb]{ 1,  .247,  .247}       3.63  & \cellcolor[rgb]{ 1,  .247,  .247}       3.54  \\
\end{tabular}%
}    		
  \label{tab:tab3bis}%
\end{table}%

\begin{table}[H]
  \centering
  \caption{Sharpe level for 90\% targeted skill using Student two-tailed test for \textbf{monthly} periods under various auto-correlation assumptions}
  \vspace{0.25cm}
  \resizebox {\textwidth} {!} {
    \begin{tabular}{|r|r|r|r|r|r|r|r|r|r|r|r|r|r|r|r|r|r|r|r|r|}
    \toprule
    \multicolumn{1}{|p{5.5em}|}{\textbf{ \textbackslash{} Observ.\newline{}Correl.}} & \textbf{6} & \textbf{12} & \textbf{18} & \textbf{24} & \textbf{30} & \textbf{36} & \textbf{42} & \textbf{48} & \textbf{54} & \textbf{60} & \textbf{66} & \textbf{72} & \textbf{78} & \textbf{84} & \textbf{90} & \textbf{96} & \textbf{102} & \textbf{108} & \textbf{114} & \textbf{120} \\
    \midrule
    \textbf{-90\%} & \cellcolor[rgb]{ 1,  .247,  .247}         9.92  & \cellcolor[rgb]{ 1,  .247,  .247}         6.25  & \cellcolor[rgb]{ 1,  .247,  .247}       4.95  & \cellcolor[rgb]{ 1,  .247,  .247}       4.22  & \cellcolor[rgb]{ 1,  .247,  .247}       3.74  & \cellcolor[rgb]{ 1,  .247,  .247}       3.40  & \cellcolor[rgb]{ 1,  .247,  .247}       3.13  & \cellcolor[rgb]{ 1,  .247,  .247}       2.92  & \cellcolor[rgb]{ 1,  .247,  .247}       2.75  & \cellcolor[rgb]{ 1,  .247,  .247}       2.60  & \cellcolor[rgb]{ 1,  .247,  .247}       2.48  & \cellcolor[rgb]{ 1,  .247,  .247}       2.37  & \cellcolor[rgb]{ 1,  .247,  .247}       2.27  & \cellcolor[rgb]{ 1,  .247,  .247}       2.19  & \cellcolor[rgb]{ 1,  .247,  .247}       2.11  & \cellcolor[rgb]{ 1,  .247,  .247}       2.05  & \cellcolor[rgb]{ 1,  .475,  .475}       1.98  & \cellcolor[rgb]{ 1,  .475,  .475}       1.93  & \cellcolor[rgb]{ 1,  .475,  .475}       1.87  & \cellcolor[rgb]{ 1,  .475,  .475}       1.83  \\
\end{tabular}%
}
  \label{tab:tab4bis}%
\end{table}%

\subsection{95\% skill target}
\subsubsection{Wald assumptions}
\vspace{-0.5cm}

\begin{table}[H]
  \centering
  \caption{Sharpe level for 95\% targeted skill using Wald two-tailed test for \textbf{daily} periods under various auto-correlation assumptions}
  \vspace{0.25cm}
  \resizebox {\textwidth} {!} {
    \begin{tabular}{|r|r|r|r|r|r|r|r|r|r|r|r|r|r|r|r|r|r|r|r|r|r|}
    \toprule
    \multicolumn{1}{|p{5.5em}|}{\textbf{ \textbackslash{} Observ.\newline{}Correl.}} & \textbf{25} & \textbf{50} & \textbf{100} & \textbf{150} & \textbf{200} & \textbf{250} & \textbf{300} & \textbf{350} & \textbf{400} & \textbf{450} & \textbf{500} & \textbf{550} & \textbf{600} & \textbf{650} & \textbf{700} & \textbf{750} & \textbf{800} & \textbf{850} & \textbf{900} & \textbf{950} & \textbf{1000} \\
    \midrule
    \textbf{-90\%} & \cellcolor[rgb]{ 1,  .247,  .247}    26.63  & \cellcolor[rgb]{ 1,  .247,  .247}     18.83  & \cellcolor[rgb]{ 1,  .247,  .247}     13.31  & \cellcolor[rgb]{ 1,  .247,  .247}     10.87  & \cellcolor[rgb]{ 1,  .247,  .247}      9.41  & \cellcolor[rgb]{ 1,  .247,  .247}     8.42  & \cellcolor[rgb]{ 1,  .247,  .247}     7.69  & \cellcolor[rgb]{ 1,  .247,  .247}      7.12  & \cellcolor[rgb]{ 1,  .247,  .247}     6.66  & \cellcolor[rgb]{ 1,  .247,  .247}     6.28  & \cellcolor[rgb]{ 1,  .247,  .247}     5.95  & \cellcolor[rgb]{ 1,  .247,  .247}     5.68  & \cellcolor[rgb]{ 1,  .247,  .247}     5.44  & \cellcolor[rgb]{ 1,  .247,  .247}     5.22  & \cellcolor[rgb]{ 1,  .247,  .247}     5.03  & \cellcolor[rgb]{ 1,  .247,  .247}     4.86  & \cellcolor[rgb]{ 1,  .247,  .247}      4.71  & \cellcolor[rgb]{ 1,  .247,  .247}     4.57  & \cellcolor[rgb]{ 1,  .247,  .247}     4.44  & \cellcolor[rgb]{ 1,  .247,  .247}     4.32  & \cellcolor[rgb]{ 1,  .247,  .247}      4.21  \\
\end{tabular}%
}
  \label{tab:tab5}%
\end{table}%

\begin{table}[H]
  \centering
  \caption{Sharpe level for 95\% targeted skill using Wald two-tailed test for \textbf{monthly} periods under various auto-correlation assumptions}
  \vspace{0.25cm}
 \resizebox {\textwidth} {!} {
    \begin{tabular}{|r|r|r|r|r|r|r|r|r|r|r|r|r|r|r|r|r|r|r|r|r|}
    \toprule
    \multicolumn{1}{|p{5.5em}|}{\textbf{ \textbackslash{} Observ.\newline{}Correl.}} & \textbf{6} & \textbf{12} & \textbf{18} & \textbf{24} & \textbf{30} & \textbf{36} & \textbf{42} & \textbf{48} & \textbf{54} & \textbf{60} & \textbf{66} & \textbf{72} & \textbf{78} & \textbf{84} & \textbf{90} & \textbf{96} & \textbf{102} & \textbf{108} & \textbf{114} & \textbf{120} \\
    \midrule
    \textbf{-90\%} & \cellcolor[rgb]{ 1,  .247,  .247}       9.65  & \cellcolor[rgb]{ 1,  .247,  .247}       6.83  & \cellcolor[rgb]{ 1,  .247,  .247}      5.57  & \cellcolor[rgb]{ 1,  .247,  .247}       4.83  & \cellcolor[rgb]{ 1,  .247,  .247}     4.32  & \cellcolor[rgb]{ 1,  .247,  .247}     3.94  & \cellcolor[rgb]{ 1,  .247,  .247}     3.65  & \cellcolor[rgb]{ 1,  .247,  .247}      3.41  & \cellcolor[rgb]{ 1,  .247,  .247}     3.22  & \cellcolor[rgb]{ 1,  .247,  .247}     3.05  & \cellcolor[rgb]{ 1,  .247,  .247}      2.91  & \cellcolor[rgb]{ 1,  .247,  .247}     2.79  & \cellcolor[rgb]{ 1,  .247,  .247}     2.68  & \cellcolor[rgb]{ 1,  .247,  .247}     2.58  & \cellcolor[rgb]{ 1,  .247,  .247}     2.49  & \cellcolor[rgb]{ 1,  .247,  .247}      2.41  & \cellcolor[rgb]{ 1,  .247,  .247}     2.34  & \cellcolor[rgb]{ 1,  .247,  .247}     2.28  & \cellcolor[rgb]{ 1,  .247,  .247}      2.21  & \cellcolor[rgb]{ 1,  .247,  .247}      2.16  \\
\end{tabular}%
}
  \label{tab:tab6}%
\end{table}%

\subsubsection{Student assumptions and one-tailed test}
\vspace{-0.5cm}

\begin{table}[H]
  \centering
  \caption{Sharpe level for 95\% targeted skill using Student one-tailed test for \textbf{daily} periods under various auto-correlation assumptions}
  \vspace{0.25cm}
  \resizebox {\textwidth} {!} {
    \begin{tabular}{|r|r|r|r|r|r|r|r|r|r|r|r|r|r|r|r|r|r|r|r|r|r|}
    \toprule
    \multicolumn{1}{|p{5.5em}|}{\textbf{ \textbackslash{} Observ.\newline{}Correl.}} & \textbf{25} & \textbf{50} & \textbf{100} & \textbf{150} & \textbf{200} & \textbf{250} & \textbf{300} & \textbf{350} & \textbf{400} & \textbf{450} & \textbf{500} & \textbf{550} & \textbf{600} & \textbf{650} & \textbf{700} & \textbf{750} & \textbf{800} & \textbf{850} & \textbf{900} & \textbf{950} & \textbf{1000} \\
    \midrule
    \textbf{-90\%} & \cellcolor[rgb]{ 1,  .247,  .247}    23.24  & \cellcolor[rgb]{ 1,  .247,  .247}       16.11  & \cellcolor[rgb]{ 1,  .247,  .247}     11.28  & \cellcolor[rgb]{ 1,  .247,  .247}        9.18  & \cellcolor[rgb]{ 1,  .247,  .247}     7.94  & \cellcolor[rgb]{ 1,  .247,  .247}     7.09  & \cellcolor[rgb]{ 1,  .247,  .247}     6.47  & \cellcolor[rgb]{ 1,  .247,  .247}     5.99  & \cellcolor[rgb]{ 1,  .247,  .247}     5.60  & \cellcolor[rgb]{ 1,  .247,  .247}     5.28  & \cellcolor[rgb]{ 1,  .247,  .247}      5.01  & \cellcolor[rgb]{ 1,  .247,  .247}     4.77  & \cellcolor[rgb]{ 1,  .247,  .247}     4.57  & \cellcolor[rgb]{ 1,  .247,  .247}     4.39  & \cellcolor[rgb]{ 1,  .247,  .247}     4.23  & \cellcolor[rgb]{ 1,  .247,  .247}     4.09  & \cellcolor[rgb]{ 1,  .247,  .247}     3.96  & \cellcolor[rgb]{ 1,  .247,  .247}     3.84  & \cellcolor[rgb]{ 1,  .247,  .247}     3.73  & \cellcolor[rgb]{ 1,  .247,  .247}     3.63  & \cellcolor[rgb]{ 1,  .247,  .247}     3.54  \\
\end{tabular}%
}
  \label{tab:tab7}%
\end{table}%

\begin{table}[H]
  \centering
  \caption{Sharpe level for 95\% targeted skill using Student one-tailed test for \textbf{monthly} periods under various auto-correlation assumptions}
  \vspace{0.25cm}
  \resizebox {\textwidth} {!} {
    \begin{tabular}{|r|r|r|r|r|r|r|r|r|r|r|r|r|r|r|r|r|r|r|r|r|}
    \toprule
    \multicolumn{1}{|p{5.5em}|}{\textbf{ \textbackslash{} Observ.\newline{}Correl.}} & \textbf{6} & \textbf{12} & \textbf{18} & \textbf{24} & \textbf{30} & \textbf{36} & \textbf{42} & \textbf{48} & \textbf{54} & \textbf{60} & \textbf{66} & \textbf{72} & \textbf{78} & \textbf{84} & \textbf{90} & \textbf{96} & \textbf{102} & \textbf{108} & \textbf{114} & \textbf{120} \\
    \midrule
    \textbf{-90\%} & \cellcolor[rgb]{ 1,  .247,  .247}       9.92  & \cellcolor[rgb]{ 1,  .247,  .247}       6.25  & \cellcolor[rgb]{ 1,  .247,  .247}      4.95  & \cellcolor[rgb]{ 1,  .247,  .247}       4.22  & \cellcolor[rgb]{ 1,  .247,  .247}     3.74  & \cellcolor[rgb]{ 1,  .247,  .247}     3.40  & \cellcolor[rgb]{ 1,  .247,  .247}      3.13  & \cellcolor[rgb]{ 1,  .247,  .247}     2.92  & \cellcolor[rgb]{ 1,  .247,  .247}     2.75  & \cellcolor[rgb]{ 1,  .247,  .247}     2.60  & \cellcolor[rgb]{ 1,  .247,  .247}     2.48  & \cellcolor[rgb]{ 1,  .247,  .247}     2.37  & \cellcolor[rgb]{ 1,  .247,  .247}     2.27  & \cellcolor[rgb]{ 1,  .247,  .247}      2.19  & \cellcolor[rgb]{ 1,  .247,  .247}       2.11  & \cellcolor[rgb]{ 1,  .247,  .247}     2.05  & \cellcolor[rgb]{ 1,  .475,  .475}      1.98  & \cellcolor[rgb]{ 1,  .475,  .475}      1.93  & \cellcolor[rgb]{ 1,  .475,  .475}      1.87  & \cellcolor[rgb]{ 1,  .475,  .475}      1.83  \\
\end{tabular}%
}
  \label{tab:tab8}%
\end{table}%

\subsubsection{Student assumptions and two-tailed test}
\vspace{-0.5cm}

\begin{table}[H]
  \centering
  \caption{Sharpe level for 95\% targeted skill using Student two-tailed test for \textbf{daily} periods under various auto-correlation assumptions}
  \vspace{0.25cm}
  \resizebox {\textwidth} {!} {
    \begin{tabular}{|r|r|r|r|r|r|r|r|r|r|r|r|r|r|r|r|r|r|r|r|r|r|}
    \toprule
    \multicolumn{1}{|p{5.5em}|}{\textbf{ \textbackslash{} Observ.\newline{}Correl.}} & \textbf{25} & \textbf{50} & \textbf{100} & \textbf{150} & \textbf{200} & \textbf{250} & \textbf{300} & \textbf{350} & \textbf{400} & \textbf{450} & \textbf{500} & \textbf{550} & \textbf{600} & \textbf{650} & \textbf{700} & \textbf{750} & \textbf{800} & \textbf{850} & \textbf{900} & \textbf{950} & \textbf{1000} \\
    \midrule
    \textbf{-90\%} & \cellcolor[rgb]{ 1,  .247,  .247}    28.04  & \cellcolor[rgb]{ 1,  .247,  .247}      19.31  & \cellcolor[rgb]{ 1,  .247,  .247}    13.48  & \cellcolor[rgb]{ 1,  .247,  .247}     10.96  & \cellcolor[rgb]{ 1,  .247,  .247}     9.47  & \cellcolor[rgb]{ 1,  .247,  .247}     8.46  & \cellcolor[rgb]{ 1,  .247,  .247}     7.72  & \cellcolor[rgb]{ 1,  .247,  .247}      7.14  & \cellcolor[rgb]{ 1,  .247,  .247}     6.68  & \cellcolor[rgb]{ 1,  .247,  .247}     6.29  & \cellcolor[rgb]{ 1,  .247,  .247}     5.97  & \cellcolor[rgb]{ 1,  .247,  .247}     5.69  & \cellcolor[rgb]{ 1,  .247,  .247}     5.45  & \cellcolor[rgb]{ 1,  .247,  .247}     5.23  & \cellcolor[rgb]{ 1,  .247,  .247}     5.04  & \cellcolor[rgb]{ 1,  .247,  .247}     4.87  & \cellcolor[rgb]{ 1,  .247,  .247}      4.71  & \cellcolor[rgb]{ 1,  .247,  .247}     4.57  & \cellcolor[rgb]{ 1,  .247,  .247}     4.44  & \cellcolor[rgb]{ 1,  .247,  .247}     4.33  & \cellcolor[rgb]{ 1,  .247,  .247}     4.22  \\
\end{tabular}%
}
  \label{tab:tab7bis}%
\end{table}%

\begin{table}[H]
  \centering
  \caption{Sharpe level for 95\% targeted skill using Student two-tailed test for \textbf{monthly} periods under various auto-correlation assumptions}
  \vspace{0.25cm}
  \resizebox {\textwidth} {!} {
    \begin{tabular}{|r|r|r|r|r|r|r|r|r|r|r|r|r|r|r|r|r|r|r|r|r|}
    \toprule
    \multicolumn{1}{|p{5.5em}|}{\textbf{ \textbackslash{} Observ.\newline{}Correl.}} & \textbf{6} & \textbf{12} & \textbf{18} & \textbf{24} & \textbf{30} & \textbf{36} & \textbf{42} & \textbf{48} & \textbf{54} & \textbf{60} & \textbf{66} & \textbf{72} & \textbf{78} & \textbf{84} & \textbf{90} & \textbf{96} & \textbf{102} & \textbf{108} & \textbf{114} & \textbf{120} \\
    \midrule
    \textbf{-90\%} & \cellcolor[rgb]{ 1,  .247,  .247}     12.66  & \cellcolor[rgb]{ 1,  .247,  .247}       7.67  & \cellcolor[rgb]{ 1,  .247,  .247}      6.00  & \cellcolor[rgb]{ 1,  .247,  .247}       5.09  & \cellcolor[rgb]{ 1,  .247,  .247}     4.50  & \cellcolor[rgb]{ 1,  .247,  .247}     4.08  & \cellcolor[rgb]{ 1,  .247,  .247}     3.76  & \cellcolor[rgb]{ 1,  .247,  .247}     3.50  & \cellcolor[rgb]{ 1,  .247,  .247}     3.29  & \cellcolor[rgb]{ 1,  .247,  .247}      3.12  & \cellcolor[rgb]{ 1,  .247,  .247}     2.97  & \cellcolor[rgb]{ 1,  .247,  .247}     2.83  & \cellcolor[rgb]{ 1,  .247,  .247}     2.72  & \cellcolor[rgb]{ 1,  .247,  .247}     2.62  & \cellcolor[rgb]{ 1,  .247,  .247}     2.53  & \cellcolor[rgb]{ 1,  .247,  .247}     2.44  & \cellcolor[rgb]{ 1,  .247,  .247}     2.37  & \cellcolor[rgb]{ 1,  .247,  .247}     2.30  & \cellcolor[rgb]{ 1,  .247,  .247}     2.24  & \cellcolor[rgb]{ 1,  .247,  .247}      2.18  \\
\end{tabular}%
}
  \label{tab:tab8bis}%
\end{table}%

\pagebreak
\subsection{Sharpe ratio significance level}
\subsubsection{Wald assumptions}
 \vspace{-0.6cm}

\begin{table}[H]
  \centering
  \caption{Sharpe for various confidence levels and number of periods for \textbf{daily} periods (with zero auto-correlation) using Wald test}
  \vspace{0.25cm}
  \resizebox {0.8 \textwidth} {!} {
    \begin{tabular}{|r|r|r|r|r|r|r|r|r|r|r|r|r|r|r|r|r|r|r|r|r|r|r|}
    \toprule
    \multicolumn{1}{|l|}{\textbf{ Skill percentage }} & \textbf{25} & \textbf{50} & \textbf{100} & \textbf{122} & \textbf{150} & \textbf{200} & \textbf{250} & \textbf{300} & \textbf{350} & \textbf{400} & \textbf{450} & \textbf{500} & \textbf{550} & \textbf{600} & \textbf{650} & \textbf{700} & \textbf{750} & \textbf{800} & \textbf{850} & \textbf{900} & \textbf{950} & \textbf{1000} \\
    \midrule
    \textbf{80\%} & \cellcolor[rgb]{ 1,  .247,  .247}                   4.07  & \cellcolor[rgb]{ 1,  .247,  .247}      2.88  & \cellcolor[rgb]{ 1,  .247,  .247}     2.03  & \cellcolor[rgb]{ 1,  .475,  .475}      1.84  & \cellcolor[rgb]{ 1,  .475,  .475}      1.66  & \cellcolor[rgb]{ 1,  .686,  .686}      1.44  & \cellcolor[rgb]{ 1,  .686,  .686}      1.29  & \cellcolor[rgb]{ 1,  .686,  .686}       1.17  & \cellcolor[rgb]{ 1,  .686,  .686}      1.09  & \cellcolor[rgb]{ 1,  .686,  .686}      1.02  & \cellcolor[rgb]{ 1,  .922,  .922}     0.96  & \cellcolor[rgb]{ 1,  .922,  .922}      0.91  & \cellcolor[rgb]{ 1,  .922,  .922}     0.87  & \cellcolor[rgb]{ 1,  .922,  .922}     0.83  & \cellcolor[rgb]{ 1,  .922,  .922}     0.80  & \cellcolor[rgb]{ 1,  .922,  .922}     0.77  & \cellcolor[rgb]{ 1,  .922,  .922}     0.74  & \cellcolor[rgb]{ 1,  .922,  .922}     0.72  & \cellcolor[rgb]{ 1,  .922,  .922}     0.70  & \cellcolor[rgb]{ 1,  .922,  .922}     0.68  & \cellcolor[rgb]{ 1,  .922,  .922}     0.66  & \cellcolor[rgb]{ 1,  .922,  .922}     0.64  \\
\end{tabular}%
  \label{tab:tab9}%
}
\end{table}%

\newpage

\begin{table}[H]
  \centering
  \caption{Sharpe for various confidence levels and number of periods for \textbf{monthly} periods (with zero auto-correlation) using Wald test}
  \vspace{0.25cm}
  \resizebox {0.8 \textwidth} {!} {
    \begin{tabular}{|r|r|r|r|r|r|r|r|r|r|r|r|r|r|r|r|r|r|r|r|r|}
    \toprule
    \multicolumn{1}{|l|}{\textbf{ Skill percentage }} & \textbf{6} & \textbf{12} & \textbf{18} & \textbf{24} & \textbf{30} & \textbf{36} & \textbf{42} & \textbf{48} & \textbf{54} & \textbf{60} & \textbf{66} & \textbf{72} & \textbf{78} & \textbf{84} & \textbf{90} & \textbf{96} & \textbf{102} & \textbf{108} & \textbf{114} & \textbf{120} \\
    \midrule
    \textbf{80\%} & \cellcolor[rgb]{ 1,  .475,  .475}                     1.81  & \cellcolor[rgb]{ 1,  .686,  .686}       1.28  & \cellcolor[rgb]{ 1,  .686,  .686}      1.05  & \cellcolor[rgb]{ 1,  .922,  .922}      0.91  & \cellcolor[rgb]{ 1,  .922,  .922}      0.81  & \cellcolor[rgb]{ 1,  .922,  .922}     0.74  & \cellcolor[rgb]{ 1,  .922,  .922}     0.69  & \cellcolor[rgb]{ 1,  .922,  .922}     0.64  & \cellcolor[rgb]{ 1,  .922,  .922}     0.60  & \cellcolor[rgb]{ 1,  .922,  .922}     0.57  & \cellcolor[rgb]{ 1,  .922,  .922}     0.55  & \cellcolor[rgb]{ 1,  .922,  .922}     0.52  & \cellcolor[rgb]{ 1,  .922,  .922}     0.50  &      0.48  &      0.47  &      0.45  &      0.44  &      0.43  &      0.42  &       0.41  \\
    \midrule
    \textbf{…} & \multicolumn{1}{l|}{  } & \multicolumn{1}{l|}{  } & \multicolumn{1}{l|}{  } & \multicolumn{1}{l|}{  } & \multicolumn{1}{l|}{  } & \multicolumn{1}{l|}{  } & \multicolumn{1}{l|}{  } & \multicolumn{1}{l|}{  } & \multicolumn{1}{l|}{  } & \multicolumn{1}{l|}{  } & \multicolumn{1}{l|}{  } & \multicolumn{1}{l|}{  } & \multicolumn{1}{l|}{  } & \multicolumn{1}{l|}{  } & \multicolumn{1}{l|}{  } & \multicolumn{1}{l|}{  } & \multicolumn{1}{l|}{  } & \multicolumn{1}{l|}{  } & \multicolumn{1}{l|}{  } & \multicolumn{1}{l|}{  } \\
\end{tabular}%
  \label{tab:tab10}%
}
\end{table}

\pagebreak
\subsubsection{Student assumptions and one-tailed test}
\vspace{-0.5cm}

\begin{table}[H]
  \centering
  \caption{Sharpe for various confidence level and number of periods for \textbf{daily} periods (with zero auto-correlation) using Student one-tailed test}
  \vspace{0.25cm}
\resizebox {0.8 \textwidth} {!} {
    \begin{tabular}{|r|r|r|r|r|r|r|r|r|r|r|r|r|r|r|r|r|r|r|r|r|r|r|}
    \toprule
    \multicolumn{1}{|l|}{\textbf{ Skill percentage }} & \textbf{25} & \textbf{50} & \textbf{100} & \textbf{122} & \textbf{150} & \textbf{200} & \textbf{250} & \textbf{300} & \textbf{350} & \textbf{400} & \textbf{450} & \textbf{500} & \textbf{550} & \textbf{600} & \textbf{650} & \textbf{700} & \textbf{750} & \textbf{800} & \textbf{850} & \textbf{900} & \textbf{950} & \textbf{1000} \\
    \midrule
    \textbf{80\%} & \cellcolor[rgb]{ 1,  .247,  .247}                   2.72  & \cellcolor[rgb]{ 1,  .475,  .475}        1.91  & \cellcolor[rgb]{ 1,  .686,  .686}      1.34  & \cellcolor[rgb]{ 1,  .686,  .686}       1.21  & \cellcolor[rgb]{ 1,  .686,  .686}      1.09  & \cellcolor[rgb]{ 1,  .922,  .922}     0.95  & \cellcolor[rgb]{ 1,  .922,  .922}     0.85  & \cellcolor[rgb]{ 1,  .922,  .922}     0.77  & \cellcolor[rgb]{ 1,  .922,  .922}     0.72  & \cellcolor[rgb]{ 1,  .922,  .922}     0.67  & \cellcolor[rgb]{ 1,  .922,  .922}     0.63  & \cellcolor[rgb]{ 1,  .922,  .922}     0.60  & \cellcolor[rgb]{ 1,  .922,  .922}     0.57  & \cellcolor[rgb]{ 1,  .922,  .922}     0.55  & \cellcolor[rgb]{ 1,  .922,  .922}     0.52  & \cellcolor[rgb]{ 1,  .922,  .922}      0.51  &      0.49  &      0.47  &      0.46  &      0.45  &      0.43  &      0.42  \\
    \midrule
    \textbf{…} & \multicolumn{1}{l|}{  } & \multicolumn{1}{l|}{  } & \multicolumn{1}{l|}{  } & \multicolumn{1}{l|}{  } & \multicolumn{1}{l|}{  } & \multicolumn{1}{l|}{  } & \multicolumn{1}{l|}{  } & \multicolumn{1}{l|}{  } & \multicolumn{1}{l|}{  } & \multicolumn{1}{l|}{  } & \multicolumn{1}{l|}{  } & \multicolumn{1}{l|}{  } & \multicolumn{1}{l|}{  } & \multicolumn{1}{l|}{  } & \multicolumn{1}{l|}{  } & \multicolumn{1}{l|}{  } & \multicolumn{1}{l|}{  } & \multicolumn{1}{l|}{  } & \multicolumn{1}{l|}{  } & \multicolumn{1}{l|}{  } & \multicolumn{1}{l|}{  } & \multicolumn{1}{l|}{  } \\
\end{tabular}%
}
  \label{tab:tab11}%
\end{table}%

\begin{table}[H]
  \centering
  \caption{Sharpe for various confidence level and number of periods for \textbf{monthly} periods (with zero auto-correlation) using Student one-tailed test}
  \vspace{0.25cm}
  \resizebox {0.80 \textwidth} {!} {
    \begin{tabular}{|r|r|r|r|r|r|r|r|r|r|r|r|r|r|r|r|r|r|r|r|r|}
    \toprule
    \multicolumn{1}{|l|}{\textbf{ Skill percentage }} & \textbf{6} & \textbf{12} & \textbf{18} & \textbf{24} & \textbf{30} & \textbf{36} & \textbf{42} & \textbf{48} & \textbf{54} & \textbf{60} & \textbf{66} & \textbf{72} & \textbf{78} & \textbf{84} & \textbf{90} & \textbf{96} & \textbf{102} & \textbf{108} & \textbf{114} & \textbf{120} \\
    \midrule
    \textbf{80\%} & \cellcolor[rgb]{ 1,  .686,  .686}                    1.30  & \cellcolor[rgb]{ 1,  .922,  .922}      0.88  & \cellcolor[rgb]{ 1,  .922,  .922}     0.70  & \cellcolor[rgb]{ 1,  .922,  .922}      0.61  & \cellcolor[rgb]{ 1,  .922,  .922}     0.54  &      0.49  &      0.45  &      0.42  &      0.40  &      0.38  &      0.36  &      0.35  &      0.33  &      0.32  &       0.31  &      0.30  &      0.29  &      0.28  &      0.27  &      0.27  \\
    \midrule
    \textbf{…} & \multicolumn{1}{l|}{  } & \multicolumn{1}{l|}{  } & \multicolumn{1}{l|}{  } & \multicolumn{1}{l|}{  } & \multicolumn{1}{l|}{  } & \multicolumn{1}{l|}{  } & \multicolumn{1}{l|}{  } & \multicolumn{1}{l|}{  } & \multicolumn{1}{l|}{  } & \multicolumn{1}{l|}{  } & \multicolumn{1}{l|}{  } & \multicolumn{1}{l|}{  } & \multicolumn{1}{l|}{  } & \multicolumn{1}{l|}{  } & \multicolumn{1}{l|}{  } & \multicolumn{1}{l|}{  } & \multicolumn{1}{l|}{  } & \multicolumn{1}{l|}{  } & \multicolumn{1}{l|}{  } & \multicolumn{1}{l|}{  } \\
\end{tabular}%
}
  \label{tab:tab12}%
\end{table}%

\subsubsection{Student assumptions and two-tailed test}
\vspace{-0.5cm}
\begin{table}[H]
  \centering
  \caption{Sharpe for various confidence level and number of periods for \textbf{daily} periods (with zero auto-correlation) using Student two-tailed test}
  \vspace{0.25cm}
  \resizebox {0.80 \textwidth} {!} {
    \begin{tabular}{|r|r|r|r|r|r|r|r|r|r|r|r|r|r|r|r|r|r|r|r|r|r|r|}
    \toprule
    \multicolumn{1}{|l|}{\textbf{ Skill percentage }} & \textbf{25} & \textbf{50} & \textbf{100} & \textbf{122} & \textbf{150} & \textbf{200} & \textbf{250} & \textbf{300} & \textbf{350} & \textbf{400} & \textbf{450} & \textbf{500} & \textbf{550} & \textbf{600} & \textbf{650} & \textbf{700} & \textbf{750} & \textbf{800} & \textbf{850} & \textbf{900} & \textbf{950} & \textbf{1000} \\
    \midrule
    \textbf{80\%} & \cellcolor[rgb]{ 1,  .247,  .247}                    4.18  & \cellcolor[rgb]{ 1,  .247,  .247}      2.92  & \cellcolor[rgb]{ 1,  .247,  .247}     2.05  & \cellcolor[rgb]{ 1,  .475,  .475}      1.85  & \cellcolor[rgb]{ 1,  .475,  .475}      1.67  & \cellcolor[rgb]{ 1,  .686,  .686}      1.44  & \cellcolor[rgb]{ 1,  .686,  .686}      1.29  & \cellcolor[rgb]{ 1,  .686,  .686}       1.18  & \cellcolor[rgb]{ 1,  .686,  .686}      1.09  & \cellcolor[rgb]{ 1,  .686,  .686}      1.02  & \cellcolor[rgb]{ 1,  .922,  .922}     0.96  & \cellcolor[rgb]{ 1,  .922,  .922}      0.91  & \cellcolor[rgb]{ 1,  .922,  .922}     0.87  & \cellcolor[rgb]{ 1,  .922,  .922}     0.83  & \cellcolor[rgb]{ 1,  .922,  .922}     0.80  & \cellcolor[rgb]{ 1,  .922,  .922}     0.77  & \cellcolor[rgb]{ 1,  .922,  .922}     0.74  & \cellcolor[rgb]{ 1,  .922,  .922}     0.72  & \cellcolor[rgb]{ 1,  .922,  .922}     0.70  & \cellcolor[rgb]{ 1,  .922,  .922}     0.68  & \cellcolor[rgb]{ 1,  .922,  .922}     0.66  & \cellcolor[rgb]{ 1,  .922,  .922}     0.64  \\
\end{tabular}%
}
  \label{tab:tab11bis}%
\end{table}%

\begin{table}[H]
  \centering
  \caption{Sharpe for various confidence level and number of periods for \textbf{monthly} periods (with zero auto-correlation) using Student two-tailed test}
  \vspace{0.25cm}
  \resizebox {0.80 \textwidth} {!} {
    \begin{tabular}{|r|r|r|r|r|r|r|r|r|r|r|r|r|r|r|r|r|r|r|r|r|}
    \toprule
    \multicolumn{1}{|l|}{\textbf{ Skill percentage }} & \textbf{6} & \textbf{12} & \textbf{18} & \textbf{24} & \textbf{30} & \textbf{36} & \textbf{42} & \textbf{48} & \textbf{54} & \textbf{60} & \textbf{66} & \textbf{72} & \textbf{78} & \textbf{84} & \textbf{90} & \textbf{96} & \textbf{102} & \textbf{108} & \textbf{114} & \textbf{120} \\
    \midrule
    \textbf{80\%} & \cellcolor[rgb]{ 1,  .247,  .247}                   2.09  & \cellcolor[rgb]{ 1,  .686,  .686}       1.36  & \cellcolor[rgb]{ 1,  .686,  .686}      1.09  & \cellcolor[rgb]{ 1,  .922,  .922}     0.93  & \cellcolor[rgb]{ 1,  .922,  .922}     0.83  & \cellcolor[rgb]{ 1,  .922,  .922}     0.75  & \cellcolor[rgb]{ 1,  .922,  .922}     0.70  & \cellcolor[rgb]{ 1,  .922,  .922}     0.65  & \cellcolor[rgb]{ 1,  .922,  .922}      0.61  & \cellcolor[rgb]{ 1,  .922,  .922}     0.58  & \cellcolor[rgb]{ 1,  .922,  .922}     0.55  & \cellcolor[rgb]{ 1,  .922,  .922}     0.53  & \cellcolor[rgb]{ 1,  .922,  .922}      0.51  &      0.49  &      0.47  &      0.46  &      0.44  &      0.43  &      0.42  &       0.41  \\
    \midrule
    \textbf{…} & \multicolumn{1}{l|}{  } & \multicolumn{1}{l|}{  } & \multicolumn{1}{l|}{  } & \multicolumn{1}{l|}{  } & \multicolumn{1}{l|}{  } & \multicolumn{1}{l|}{  } & \multicolumn{1}{l|}{  } & \multicolumn{1}{l|}{  } & \multicolumn{1}{l|}{  } & \multicolumn{1}{l|}{  } & \multicolumn{1}{l|}{  } & \multicolumn{1}{l|}{  } & \multicolumn{1}{l|}{  } & \multicolumn{1}{l|}{  } & \multicolumn{1}{l|}{  } & \multicolumn{1}{l|}{  } & \multicolumn{1}{l|}{  } & \multicolumn{1}{l|}{  } & \multicolumn{1}{l|}{  } & \multicolumn{1}{l|}{  } \\
\end{tabular}%
}
  \label{tab:tab12bis}%
\end{table}%

\pagebreak
\subsection{Percentage of skill for a given level of Sharpe ratio}
\subsubsection{Sharpe ratio of 0.5}
\vspace{-0.5cm}

\begin{table}[H]
  \centering
  \caption{Skill percentage for a Sharpe ratio of 0.5 for \textbf{daily} periods under various auto-correlation assumptions under Wald assumptions (two-tailed test)}
  \vspace{0.25cm}
  \resizebox {0.85 \textwidth} {!} {
    \begin{tabular}{|r|r|r|r|r|r|r|r|r|r|r|r|r|r|r|r|r|r|r|r|r|r|}
    \toprule
    \multicolumn{1}{|p{5.5em}|}{\textbf{ \textbackslash{} Observ.\newline{}Correl.}} & \textbf{25} & \textbf{50} & \textbf{100} & \textbf{150} & \textbf{200} & \textbf{250} & \textbf{300} & \textbf{350} & \textbf{400} & \textbf{450} & \textbf{500} & \textbf{550} & \textbf{600} & \textbf{650} & \textbf{700} & \textbf{750} & \textbf{800} & \textbf{850} & \textbf{900} & \textbf{950} & \textbf{1000} \\
    \midrule
    \textbf{-90\%} & 3\%   & 4\%   & 6\%   & 7\%   & 8\%   & 9\%   & 10\%  & 11\%  & 12\%  & 12\%  & 13\%  & 14\%  & 14\%  & 15\%  & 15\%  & 16\%  & 16\%  & 17\%  & 17\%  & 18\%  & 18\% \\
    \midrule
    \textbf{-80\%} & 4\%   & 6\%   & 8\%   & 10\%  & 12\%  & 13\%  & 15\%  & 16\%  & 17\%  & 18\%  & 19\%  & 20\%  & 20\%  & 21\%  & 22\%  & 23\%  & 24\%  & 24\%  & 25\%  & 26\%  & 26\% \\
    \midrule
    \textbf{-70\%} & 5\%   & 7\%   & 11\%  & 13\%  & 15\%  & 17\%  & 18\%  & 20\%  & 21\%  & 22\%  & 23\%  & 24\%  & 26\%  & 27\%  & 28\%  & 28\%  & 29\%  & 30\%  & 31\%  & 32\%  & 33\% \\
    \midrule
    \textbf{-60\%} & 6\%   & 9\%   & 13\%  & 15\%  & 18\%  & 20\%  & 22\%  & 23\%  & 25\%  & 26\%  & 28\%  & 29\%  & 30\%  & 31\%  & 32\%  & 33\%  & 35\%  & 36\%  & 36\%  & 37\%  & 38\% \\
    \midrule
    \textbf{-50\%} & 7\%   & 10\%  & 14\%  & 18\%  & 20\%  & 23\%  & 25\%  & 27\%  & 28\%  & 30\%  & 32\%  & 33\%  & 34\%  & 36\%  & 37\%  & 38\%  & 39\%  & 40\%  & 42\%  & 43\%  & 44\% \\
    \midrule
    \textbf{-40\%} & 8\%   & 12\%  & 16\%  & 20\%  & 23\%  & 26\%  & 28\%  & 30\%  & 32\%  & 34\%  & 36\%  & 37\%  & 39\%  & 40\%  & 42\%  & 43\%  & 44\%  & 45\%  & 46\%  & 48\%  & 49\% \\
    \midrule
    \textbf{-30\%} & 9\%   & 13\%  & 18\%  & 22\%  & 26\%  & 29\%  & 31\%  & 33\%  & 36\%  & 38\%  & 40\%  & 41\%  & 43\%  & 44\%  & 46\%  & 47\%  & 49\%  & 50\%  & 51\%  & 52\%  & 54\% \\
\end{tabular}%
}
  \label{tab:tab13}%
\end{table}%

\begin{table}[H]
  \centering
  \caption{Skill percentage for a Sharpe ratio of 0.5 for \textbf{monthly} periods under various auto-correlation assumptions under Wald assumptions (two-tailed test)}
  \vspace{0.25cm}
  \resizebox {0.85 \textwidth} {!} {
    \begin{tabular}{|r|r|r|r|r|r|r|r|r|r|r|r|r|r|r|r|r|r|r|r|r|}
    \toprule
    \multicolumn{1}{|p{5.5em}|}{\textbf{ \textbackslash{} Observ.\newline{}Correl.}} & \textbf{6} & \textbf{12} & \textbf{18} & \textbf{24} & \textbf{30} & \textbf{36} & \textbf{42} & \textbf{48} & \textbf{54} & \textbf{60} & \textbf{66} & \textbf{72} & \textbf{78} & \textbf{84} & \textbf{90} & \textbf{96} & \textbf{102} & \textbf{108} & \textbf{114} & \textbf{120} \\
    \midrule
    \textbf{-90\%} & 8\%   & 11\%  & 14\%  & 16\%  & 18\%  & 20\%  & 21\%  & 23\%  & 24\%  & 25\%  & 26\%  & 27\%  & 29\%  & 30\%  & 31\%  & 32\%  & 32\%  & 33\%  & 34\%  & 35\% \\
    \midrule
    \textbf{-80\%} & 11\%  & 15\%  & 19\%  & 22\%  & 24\%  & 26\%  & 28\%  & 30\%  & 32\%  & 33\%  & 35\%  & 36\%  & 38\%  & 39\%  & 40\%  & 42\%  & 43\%  & 44\%  & 45\%  & 46\% \\
    \midrule
    \textbf{-70\%} & 13\%  & 18\%  & 22\%  & 26\%  & 29\%  & 31\%  & 34\%  & 36\%  & 38\%  & 40\%  & 41\%  & 43\%  & 45\%  & 46\%  & 48\%  & 49\%  & 50\%  & 51\%  & 53\%  & 54\% \\
    \midrule
    \textbf{-60\%} & 15\%  & 21\%  & 26\%  & 30\%  & 33\%  & 36\%  & 38\%  & 41\%  & 43\%  & 45\%  & 47\%  & 49\%  & 51\%  & 52\%  & 54\%  & 55\%  & 57\%  & 58\%  & 59\%  & 60\% \\
    \midrule
    \textbf{-50\%} & 17\%  & 24\%  & 29\%  & 33\%  & 37\%  & 40\%  & 43\%  & 46\%  & 48\%  & 50\%  & 52\%  & 54\%  & 56\%  & 58\%  & 60\%  & 61\%  & 62\%  & 64\%  & 65\%  & 66\% \\
    \midrule
    \textbf{-40\%} & 19\%  & 27\%  & 32\%  & 37\%  & 41\%  & 44\%  & 48\%  & 50\%  & 53\%  & 55\%  & 57\%  & 60\%  & 61\%  & 63\%  & 65\%  & 66\%  & 68\%  & 69\%  & 71\%  & 72\% \\
    \midrule
    \textbf{-30\%} & 21\%  & 29\%  & 36\%  & 41\%  & 45\%  & 49\%  & 52\%  & 55\%  & 58\%  & 60\%  & 62\%  & 64\%  & 66\%  & 68\%  & 70\%  & 71\%  & 73\%  & 74\%  & 75\%  & 77\% \\
\end{tabular}%
}
  \label{tab:tab14}%
\end{table}%

\pagebreak

\begin{table}[H]
  \centering
  \caption{Skill percentage for a Sharpe ratio of 0.5 for \textbf{daily} periods under various auto-correlation assumptions under Student assumptions (one-tailed test)}
  \vspace{0.25cm}
  \resizebox {0.88 \textwidth} {!} {
    \begin{tabular}{|r|r|r|r|r|r|r|r|r|r|r|r|r|r|r|r|r|r|r|r|r|r|}
    \toprule
    \multicolumn{1}{|p{5.5em}|}{\textbf{ \textbackslash{} Observ.\newline{}Correl.}} & \textbf{25} & \textbf{50} & \textbf{100} & \textbf{150} & \textbf{200} & \textbf{250} & \textbf{300} & \textbf{350} & \textbf{400} & \textbf{450} & \textbf{500} & \textbf{550} & \textbf{600} & \textbf{650} & \textbf{700} & \textbf{750} & \textbf{800} & \textbf{850} & \textbf{900} & \textbf{950} & \textbf{1000} \\
    \midrule
    \textbf{-90\%} & 51\%  & 52\%  & 53\%  & 54\%  & 54\%  & 55\%  & 55\%  & 55\%  & 56\%  & 56\%  & 57\%  & 57\%  & 57\%  & 57\%  & 58\%  & 58\%  & 58\%  & 58\%  & 59\%  & 59\%  & 59\% \\
    \midrule
    \textbf{-80\%} & 52\%  & 53\%  & 54\%  & 55\%  & 56\%  & 57\%  & 57\%  & 58\%  & 58\%  & 59\%  & 59\%  & 60\%  & 60\%  & 61\%  & 61\%  & 61\%  & 62\%  & 62\%  & 62\%  & 63\%  & 63\% \\
    \midrule
    \textbf{-70\%} & 53\%  & 54\%  & 55\%  & 56\%  & 57\%  & 58\%  & 59\%  & 60\%  & 60\%  & 61\%  & 62\%  & 62\%  & 63\%  & 63\%  & 64\%  & 64\%  & 65\%  & 65\%  & 66\%  & 66\%  & 66\% \\
    \midrule
    \textbf{-60\%} & 53\%  & 54\%  & 56\%  & 58\%  & 59\%  & 60\%  & 61\%  & 62\%  & 62\%  & 63\%  & 64\%  & 64\%  & 65\%  & 66\%  & 66\%  & 67\%  & 67\%  & 68\%  & 68\%  & 69\%  & 69\% \\
    \midrule
    \textbf{-50\%} & 54\%  & 55\%  & 57\%  & 59\%  & 60\%  & 61\%  & 62\%  & 63\%  & 64\%  & 65\%  & 66\%  & 67\%  & 67\%  & 68\%  & 69\%  & 69\%  & 70\%  & 70\%  & 71\%  & 71\%  & 72\% \\
    \midrule
    \textbf{-40\%} & 54\%  & 56\%  & 58\%  & 60\%  & 61\%  & 63\%  & 64\%  & 65\%  & 66\%  & 67\%  & 68\%  & 69\%  & 69\%  & 70\%  & 71\%  & 71\%  & 72\%  & 73\%  & 73\%  & 74\%  & 74\% \\
    \midrule
    \textbf{-30\%} & 55\%  & 56\%  & 59\%  & 61\%  & 63\%  & 64\%  & 66\%  & 67\%  & 68\%  & 69\%  & 70\%  & 71\%  & 71\%  & 72\%  & 73\%  & 74\%  & 74\%  & 75\%  & 76\%  & 76\%  & 77\% \\
\end{tabular}%
}
  \label{tab:tab15}%
\end{table}%

\pagebreak
\begin{table}[H]
  \centering
  \caption{Skill percentage for a Sharpe ratio of 0.5 for \textbf{monthly} periods under various auto-correlation assumptions under Student assumptions (one-tailed test)}
  \vspace{0.25cm}
  \resizebox {0.85 \textwidth} {!} {
    \begin{tabular}{|r|r|r|r|r|r|r|r|r|r|r|r|r|r|r|r|r|r|r|r|r|}
    \toprule
    \multicolumn{1}{|p{5.5em}|}{\textbf{ \textbackslash{} Observ.\newline{}Correl.}} & \textbf{6} & \textbf{12} & \textbf{18} & \textbf{24} & \textbf{30} & \textbf{36} & \textbf{42} & \textbf{48} & \textbf{54} & \textbf{60} & \textbf{66} & \textbf{72} & \textbf{78} & \textbf{84} & \textbf{90} & \textbf{96} & \textbf{102} & \textbf{108} & \textbf{114} & \textbf{120} \\
    \midrule
    \textbf{-90\%} & 54\%  & 56\%  & 57\%  & 58\%  & 59\%  & 60\%  & 61\%  & 61\%  & 62\%  & 63\%  & 63\%  & 64\%  & 64\%  & 65\%  & 65\%  & 66\%  & 66\%  & 67\%  & 67\%  & 67\% \\
    \midrule
    \textbf{-80\%} & 55\%  & 57\%  & 59\%  & 61\%  & 62\%  & 63\%  & 64\%  & 65\%  & 66\%  & 67\%  & 67\%  & 68\%  & 69\%  & 69\%  & 70\%  & 71\%  & 71\%  & 72\%  & 72\%  & 73\% \\
    \midrule
    \textbf{-70\%} & 56\%  & 59\%  & 61\%  & 63\%  & 64\%  & 66\%  & 67\%  & 68\%  & 69\%  & 70\%  & 71\%  & 71\%  & 72\%  & 73\%  & 74\%  & 74\%  & 75\%  & 76\%  & 76\%  & 77\% \\
    \midrule
    \textbf{-60\%} & 57\%  & 60\%  & 63\%  & 65\%  & 66\%  & 68\%  & 69\%  & 70\%  & 71\%  & 72\%  & 73\%  & 74\%  & 75\%  & 76\%  & 77\%  & 78\%  & 78\%  & 79\%  & 80\%  & \cellcolor[rgb]{ .788,  1,  .882}80\% \\
    \midrule
    \textbf{-50\%} & 58\%  & 62\%  & 64\%  & 66\%  & 68\%  & 70\%  & 71\%  & 73\%  & 74\%  & 75\%  & 76\%  & 77\%  & 78\%  & 79\%  & 80\%  & \cellcolor[rgb]{ .788,  1,  .882}80\% & \cellcolor[rgb]{ .788,  1,  .882}81\% & \cellcolor[rgb]{ .788,  1,  .882}82\% & \cellcolor[rgb]{ .788,  1,  .882}82\% & \cellcolor[rgb]{ .788,  1,  .882}83\% \\
\end{tabular}%
}
  \label{tab:tab16}%
\end{table}%

\begin{table}[H]
  \centering
  \caption{Skill percentage for a Sharpe ratio of 0.5 for \textbf{daily} periods under various auto-correlation assumptions under Student assumptions (two-tailed test)}
  \vspace{0.25cm}
  \resizebox {0.85 \textwidth} {!} {
    \begin{tabular}{|r|r|r|r|r|r|r|r|r|r|r|r|r|r|r|r|r|r|r|r|r|r|}
    \toprule
    \multicolumn{1}{|p{5.5em}|}{\textbf{ \textbackslash{} Observ.\newline{}Correl.}} & \textbf{25} & \textbf{50} & \textbf{100} & \textbf{150} & \textbf{200} & \textbf{250} & \textbf{300} & \textbf{350} & \textbf{400} & \textbf{450} & \textbf{500} & \textbf{550} & \textbf{600} & \textbf{650} & \textbf{700} & \textbf{750} & \textbf{800} & \textbf{850} & \textbf{900} & \textbf{950} & \textbf{1000} \\
    \midrule
    \textbf{-90\%} & 3\%   & 4\%   & 6\%   & 7\%   & 8\%   & 9\%   & 10\%  & 11\%  & 12\%  & 12\%  & 13\%  & 14\%  & 14\%  & 15\%  & 15\%  & 16\%  & 16\%  & 17\%  & 17\%  & 18\%  & 18\% \\
    \midrule
    \textbf{-80\%} & 4\%   & 6\%   & 8\%   & 10\%  & 12\%  & 13\%  & 15\%  & 16\%  & 17\%  & 18\%  & 19\%  & 20\%  & 20\%  & 21\%  & 22\%  & 23\%  & 24\%  & 24\%  & 25\%  & 26\%  & 26\% \\
    \midrule
    \textbf{-70\%} & 5\%   & 7\%   & 11\%  & 13\%  & 15\%  & 17\%  & 18\%  & 20\%  & 21\%  & 22\%  & 23\%  & 24\%  & 26\%  & 27\%  & 28\%  & 28\%  & 29\%  & 30\%  & 31\%  & 32\%  & 33\% \\
    \midrule
    \textbf{-60\%} & 6\%   & 9\%   & 13\%  & 15\%  & 18\%  & 20\%  & 22\%  & 23\%  & 25\%  & 26\%  & 28\%  & 29\%  & 30\%  & 31\%  & 32\%  & 33\%  & 35\%  & 35\%  & 36\%  & 37\%  & 38\% \\
    \midrule
    \textbf{-50\%} & 7\%   & 10\%  & 14\%  & 18\%  & 20\%  & 23\%  & 25\%  & 27\%  & 28\%  & 30\%  & 32\%  & 33\%  & 34\%  & 36\%  & 37\%  & 38\%  & 39\%  & 40\%  & 42\%  & 43\%  & 44\% \\
    \midrule
    \textbf{-40\%} & 8\%   & 12\%  & 16\%  & 20\%  & 23\%  & 26\%  & 28\%  & 30\%  & 32\%  & 34\%  & 36\%  & 37\%  & 39\%  & 40\%  & 42\%  & 43\%  & 44\%  & 45\%  & 46\%  & 48\%  & 49\% \\
    \midrule
    \textbf{-30\%} & 9\%   & 13\%  & 18\%  & 22\%  & 26\%  & 29\%  & 31\%  & 33\%  & 36\%  & 38\%  & 39\%  & 41\%  & 43\%  & 44\%  & 46\%  & 47\%  & 49\%  & 50\%  & 51\%  & 52\%  & 54\% \\
\end{tabular}%
}
  \label{tab:tab15bis}%
\end{table}%

\begin{table}[H]
  \centering
  \caption{Skill percentage for a Sharpe ratio of 0.5 for \textbf{monthly} periods under various auto-correlation assumptions under Student assumptions (two-tailed test)}
  \vspace{0.25cm}
  \resizebox {0.85 \textwidth} {!} {
    \begin{tabular}{|r|r|r|r|r|r|r|r|r|r|r|r|r|r|r|r|r|r|r|r|r|}
    \toprule
    \multicolumn{1}{|p{5.5em}|}{\textbf{ \textbackslash{} Observ.\newline{}Correl.}} & \textbf{6} & \textbf{12} & \textbf{18} & \textbf{24} & \textbf{30} & \textbf{36} & \textbf{42} & \textbf{48} & \textbf{54} & \textbf{60} & \textbf{66} & \textbf{72} & \textbf{78} & \textbf{84} & \textbf{90} & \textbf{96} & \textbf{102} & \textbf{108} & \textbf{114} & \textbf{120} \\
    \midrule
    \textbf{-90\%} & 8\%   & 11\%  & 14\%  & 16\%  & 18\%  & 19\%  & 21\%  & 22\%  & 24\%  & 25\%  & 26\%  & 27\%  & 28\%  & 29\%  & 30\%  & 31\%  & 32\%  & 33\%  & 34\%  & 35\% \\
    \midrule
    \textbf{-80\%} & 10\%  & 15\%  & 18\%  & 21\%  & 24\%  & 26\%  & 28\%  & 30\%  & 32\%  & 33\%  & 35\%  & 36\%  & 38\%  & 39\%  & 40\%  & 41\%  & 43\%  & 44\%  & 45\%  & 46\% \\
    \midrule
    \textbf{-70\%} & 12\%  & 18\%  & 22\%  & 25\%  & 28\%  & 31\%  & 33\%  & 36\%  & 38\%  & 39\%  & 41\%  & 43\%  & 44\%  & 46\%  & 47\%  & 49\%  & 50\%  & 51\%  & 52\%  & 54\% \\
    \midrule
    \textbf{-60\%} & 14\%  & 21\%  & 25\%  & 29\%  & 33\%  & 36\%  & 38\%  & 41\%  & 43\%  & 45\%  & 47\%  & 49\%  & 50\%  & 52\%  & 54\%  & 55\%  & 56\%  & 58\%  & 59\%  & 60\% \\
    \midrule
    \textbf{-50\%} & 16\%  & 23\%  & 29\%  & 33\%  & 37\%  & 40\%  & 43\%  & 45\%  & 48\%  & 50\%  & 52\%  & 54\%  & 56\%  & 58\%  & 59\%  & 61\%  & 62\%  & 64\%  & 65\%  & 66\% \\
    \midrule
    \textbf{-40\%} & 18\%  & 26\%  & 32\%  & 36\%  & 41\%  & 44\%  & 47\%  & 50\%  & 53\%  & 55\%  & 57\%  & 59\%  & 61\%  & 63\%  & 65\%  & 66\%  & 68\%  & 69\%  & 70\%  & 72\% \\
    \midrule
    \textbf{-30\%} & 20\%  & 29\%  & 35\%  & 40\%  & 44\%  & 48\%  & 52\%  & 55\%  & 57\%  & 60\%  & 62\%  & 64\%  & 66\%  & 68\%  & 70\%  & 71\%  & 73\%  & 74\%  & 75\%  & 76\% \\
\end{tabular}%
  \label{tab:tab16bis}%
}
\end{table}%

\pagebreak
\subsubsection{Sharpe ratio of 1.0}
\vspace{-0.5cm}

\begin{table}[H]
  \centering
  \caption{Skill percentage for a Sharpe ratio of 1.0 for \textbf{daily} periods under various auto-correlation assumptions under Wald assumptions (two-tailed test)}
  \vspace{0.25cm}
  \resizebox {0.8 \textwidth} {!} {
    \begin{tabular}{|r|r|r|r|r|r|r|r|r|r|r|r|r|r|r|r|r|r|r|r|r|r|}
    \toprule
    \multicolumn{1}{|p{5.5em}|}{\textbf{ \textbackslash{} Observ.\newline{}Correl.}} & \textbf{25} & \textbf{50} & \textbf{100} & \textbf{150} & \textbf{200} & \textbf{250} & \textbf{300} & \textbf{350} & \textbf{400} & \textbf{450} & \textbf{500} & \textbf{550} & \textbf{600} & \textbf{650} & \textbf{700} & \textbf{750} & \textbf{800} & \textbf{850} & \textbf{900} & \textbf{950} & \textbf{1000} \\
    \midrule
    \textbf{-90\%} & 6\%   & 8\%   & 12\%  & 14\%  & 16\%  & 18\%  & 20\%  & 22\%  & 23\%  & 25\%  & 26\%  & 27\%  & 28\%  & 29\%  & 30\%  & 31\%  & 32\%  & 33\%  & 34\%  & 35\%  & 36\% \\
    \midrule
    \textbf{-80\%} & 8\%   & 12\%  & 17\%  & 20\%  & 24\%  & 26\%  & 29\%  & 31\%  & 33\%  & 35\%  & 36\%  & 38\%  & 40\%  & 41\%  & 42\%  & 44\%  & 45\%  & 46\%  & 47\%  & 49\%  & 50\% \\
    \midrule
    \textbf{-70\%} & 11\%  & 15\%  & 21\%  & 26\%  & 29\%  & 33\%  & 36\%  & 38\%  & 41\%  & 43\%  & 45\%  & 47\%  & 49\%  & 50\%  & 52\%  & 53\%  & 55\%  & 56\%  & 58\%  & 59\%  & 60\% \\
    \midrule
    \textbf{-60\%} & 13\%  & 18\%  & 25\%  & 30\%  & 35\%  & 38\%  & 42\%  & 45\%  & 47\%  & 50\%  & 52\%  & 54\%  & 56\%  & 58\%  & 60\%  & 61\%  & 63\%  & 64\%  & 66\%  & 67\%  & 68\% \\
    \midrule
    \textbf{-50\%} & 14\%  & 20\%  & 28\%  & 34\%  & 39\%  & 44\%  & 47\%  & 50\%  & 53\%  & 56\%  & 59\%  & 61\%  & 63\%  & 65\%  & 67\%  & 68\%  & 70\%  & 71\%  & 73\%  & 74\%  & 75\% \\
    \midrule
    \textbf{-40\%} & 16\%  & 23\%  & 32\%  & 39\%  & 44\%  & 49\%  & 53\%  & 56\%  & 59\%  & 62\%  & 64\%  & 67\%  & 69\%  & 71\%  & 73\%  & 74\%  & 76\%  & 77\%  & 78\%  & 80\%  & \cellcolor[rgb]{ .788,  1,  .882}81\% \\
\end{tabular}%
}
  \label{tab:tab17}%
\end{table}%

\pagebreak

\begin{table}[H]
  \centering
  \caption{Skill percentage for a Sharpe ratio of 1.0 for \textbf{monthly} periods under various auto-correlation assumptions under Wald assumptions (two-tailed test)}
  \vspace{0.25cm}
  \resizebox {0.8 \textwidth} {!} {
    \begin{tabular}{|r|r|r|r|r|r|r|r|r|r|r|r|r|r|r|r|r|r|r|r|r|}
    \toprule
    \multicolumn{1}{|p{5.5em}|}{\textbf{ \textbackslash{} Observ.\newline{}Correl.}} & \textbf{6} & \textbf{12} & \textbf{18} & \textbf{24} & \textbf{30} & \textbf{36} & \textbf{42} & \textbf{48} & \textbf{54} & \textbf{60} & \textbf{66} & \textbf{72} & \textbf{78} & \textbf{84} & \textbf{90} & \textbf{96} & \textbf{102} & \textbf{108} & \textbf{114} & \textbf{120} \\
    \midrule
    \textbf{-90\%} & 16\%  & 23\%  & 27\%  & 32\%  & 35\%  & 38\%  & 41\%  & 43\%  & 46\%  & 48\%  & 50\%  & 52\%  & 54\%  & 55\%  & 57\%  & 58\%  & 60\%  & 61\%  & 62\%  & 64\% \\
    \midrule
    \textbf{-80\%} & 22\%  & 30\%  & 36\%  & 42\%  & 46\%  & 50\%  & 53\%  & 56\%  & 59\%  & 61\%  & 64\%  & 66\%  & 68\%  & 69\%  & 71\%  & 73\%  & 74\%  & 75\%  & 77\%  & 78\% \\
    \midrule
    \textbf{-70\%} & 26\%  & 36\%  & 43\%  & 49\%  & 54\%  & 58\%  & 62\%  & 65\%  & 68\%  & 70\%  & 72\%  & 75\%  & 76\%  & 78\%  & 80\%  & \cellcolor[rgb]{ .788,  1,  .882}81\% & \cellcolor[rgb]{ .788,  1,  .882}82\% & \cellcolor[rgb]{ .788,  1,  .882}84\% & \cellcolor[rgb]{ .788,  1,  .882}85\% & \cellcolor[rgb]{ .788,  1,  .882}86\% \\
    \midrule
    \textbf{-60\%} & 30\%  & 41\%  & 49\%  & 55\%  & 60\%  & 65\%  & 69\%  & 72\%  & 75\%  & 77\%  & 79\%  & \cellcolor[rgb]{ .788,  1,  .882}81\% & \cellcolor[rgb]{ .788,  1,  .882}83\% & \cellcolor[rgb]{ .788,  1,  .882}85\% & \cellcolor[rgb]{ .788,  1,  .882}86\% & \cellcolor[rgb]{ .788,  1,  .882}87\% & \cellcolor[rgb]{ .788,  1,  .882}88\% & \cellcolor[rgb]{ .788,  1,  .882}89\% & \cellcolor[rgb]{ .592,  1,  .776}90\% & \cellcolor[rgb]{ .592,  1,  .776}91\% \\
\end{tabular}%
  \label{tab:tab18}%
}
\end{table}%

\begin{table}[H]
  \centering
  \caption{Skill percentage for a Sharpe ratio of 1.0 for \textbf{daily} periods under various auto-correlation assumptions under Student assumptions (one-tailed test)}
  \vspace{0.25cm}
  \resizebox {0.8 \textwidth} {!} {
    \begin{tabular}{|r|r|r|r|r|r|r|r|r|r|r|r|r|r|r|r|r|r|r|r|r|r|}
    \toprule
    \multicolumn{1}{|p{5.5em}|}{\textbf{ \textbackslash{} Observ.\newline{}Correl.}} & \textbf{25} & \textbf{50} & \textbf{100} & \textbf{150} & \textbf{200} & \textbf{250} & \textbf{300} & \textbf{350} & \textbf{400} & \textbf{450} & \textbf{500} & \textbf{550} & \textbf{600} & \textbf{650} & \textbf{700} & \textbf{750} & \textbf{800} & \textbf{850} & \textbf{900} & \textbf{950} & \textbf{1000} \\
    \midrule
    \textbf{-90\%} & 53\%  & 54\%  & 56\%  & 57\%  & 58\%  & 59\%  & 60\%  & 61\%  & 62\%  & 62\%  & 63\%  & 63\%  & 64\%  & 65\%  & 65\%  & 66\%  & 66\%  & 67\%  & 67\%  & 67\%  & 68\% \\
    \midrule
    \textbf{-80\%} & 54\%  & 56\%  & 58\%  & 60\%  & 62\%  & 63\%  & 64\%  & 65\%  & 66\%  & 67\%  & 68\%  & 69\%  & 70\%  & 71\%  & 71\%  & 72\%  & 73\%  & 73\%  & 74\%  & 74\%  & 75\% \\
    \midrule
    \textbf{-70\%} & 55\%  & 57\%  & 60\%  & 63\%  & 65\%  & 66\%  & 68\%  & 69\%  & 70\%  & 71\%  & 72\%  & 73\%  & 74\%  & 75\%  & 76\%  & 77\%  & 77\%  & 78\%  & 79\%  & 79\%  & 80\% \\
    \midrule
    \textbf{-60\%} & 56\%  & 59\%  & 62\%  & 65\%  & 67\%  & 69\%  & 71\%  & 72\%  & 74\%  & 75\%  & 76\%  & 77\%  & 78\%  & 79\%  & 80\%  & \cellcolor[rgb]{ .788,  1,  .882}81\% & \cellcolor[rgb]{ .788,  1,  .882}81\% & \cellcolor[rgb]{ .788,  1,  .882}82\% & \cellcolor[rgb]{ .788,  1,  .882}83\% & \cellcolor[rgb]{ .788,  1,  .882}83\% & \cellcolor[rgb]{ .788,  1,  .882}84\% \\
    \midrule
    \textbf{-50\%} & 57\%  & 60\%  & 64\%  & 67\%  & 70\%  & 72\%  & 74\%  & 75\%  & 77\%  & 78\%  & 79\%  & \cellcolor[rgb]{ .788,  1,  .882}80\% & \cellcolor[rgb]{ .788,  1,  .882}81\% & \cellcolor[rgb]{ .788,  1,  .882}82\% & \cellcolor[rgb]{ .788,  1,  .882}83\% & \cellcolor[rgb]{ .788,  1,  .882}84\% & \cellcolor[rgb]{ .788,  1,  .882}85\% & \cellcolor[rgb]{ .788,  1,  .882}86\% & \cellcolor[rgb]{ .788,  1,  .882}86\% & \cellcolor[rgb]{ .788,  1,  .882}87\% & \cellcolor[rgb]{ .788,  1,  .882}88\% \\
\end{tabular}%
}
  \label{tab:tab19}%
\end{table}%

\begin{table}[H]
  \centering
  \caption{Skill percentage for a Sharpe ratio of 1.0 for \textbf{monthly} periods under various auto-correlation assumptions under Student assumptions (one-tailed test)}
  \vspace{0.25cm}
  \resizebox {0.8 \textwidth} {!} {
    \begin{tabular}{|r|r|r|r|r|r|r|r|r|r|r|r|r|r|r|r|r|r|r|r|r|}
    \toprule
    \multicolumn{1}{|p{5.5em}|}{\textbf{ \textbackslash{} Observ.\newline{}Correl.}} & \textbf{6} & \textbf{12} & \textbf{18} & \textbf{24} & \textbf{30} & \textbf{36} & \textbf{42} & \textbf{48} & \textbf{54} & \textbf{60} & \textbf{66} & \textbf{72} & \textbf{78} & \textbf{84} & \textbf{90} & \textbf{96} & \textbf{102} & \textbf{108} & \textbf{114} & \textbf{120} \\
    \midrule
    \textbf{-90\%} & 58\%  & 61\%  & 64\%  & 66\%  & 67\%  & 69\%  & 70\%  & 72\%  & 73\%  & 74\%  & 75\%  & 76\%  & 77\%  & 78\%  & 78\%  & 79\%  & 80\%  & 80\%  & 81\%  & 82\% \\
    \midrule
    \textbf{-80\%} & 60\%  & 65\%  & 68\%  & 71\%  & 73\%  & 75\%  & 76\%  & 78\%  & 79\%  & \cellcolor[rgb]{ .788,  1,  .882}80\% & \cellcolor[rgb]{ .788,  1,  .882}82\% & \cellcolor[rgb]{ .788,  1,  .882}83\% & \cellcolor[rgb]{ .788,  1,  .882}84\% & \cellcolor[rgb]{ .788,  1,  .882}85\% & \cellcolor[rgb]{ .788,  1,  .882}85\% & \cellcolor[rgb]{ .788,  1,  .882}86\% & \cellcolor[rgb]{ .788,  1,  .882}87\% & \cellcolor[rgb]{ .788,  1,  .882}88\% & \cellcolor[rgb]{ .788,  1,  .882}88\% & \cellcolor[rgb]{ .788,  1,  .882}89\% \\
    \midrule
    \textbf{-70\%} & 62\%  & 67\%  & 71\%  & 74\%  & 77\%  & 79\%  & \cellcolor[rgb]{ .788,  1,  .882}81\% & \cellcolor[rgb]{ .788,  1,  .882}82\% & \cellcolor[rgb]{ .788,  1,  .882}84\% & \cellcolor[rgb]{ .788,  1,  .882}85\% & \cellcolor[rgb]{ .788,  1,  .882}86\% & \cellcolor[rgb]{ .788,  1,  .882}87\% & \cellcolor[rgb]{ .788,  1,  .882}88\% & \cellcolor[rgb]{ .788,  1,  .882}89\% & \cellcolor[rgb]{ .788,  1,  .882}90\% & \cellcolor[rgb]{ .592,  1,  .776}90\% & \cellcolor[rgb]{ .592,  1,  .776}91\% & \cellcolor[rgb]{ .592,  1,  .776}92\% & \cellcolor[rgb]{ .592,  1,  .776}92\% & \cellcolor[rgb]{ .592,  1,  .776}93\% \\
\end{tabular}%
}
  \label{tab:tab20}%
\end{table}%

\begin{table}[H]
  \centering
  \caption{Skill percentage for a Sharpe ratio of 1.0 for \textbf{daily} periods under various auto-correlation assumptions under Student assumptions (two-tailed test)}
  \vspace{0.25cm}
  \resizebox {0.8 \textwidth} {!} {
    \begin{tabular}{|r|r|r|r|r|r|r|r|r|r|r|r|r|r|r|r|r|r|r|r|r|r|}
    \toprule
    \multicolumn{1}{|p{5.5em}|}{\textbf{ \textbackslash{} Observ.\newline{}Correl.}} & \textbf{25} & \textbf{50} & \textbf{100} & \textbf{150} & \textbf{200} & \textbf{250} & \textbf{300} & \textbf{350} & \textbf{400} & \textbf{450} & \textbf{500} & \textbf{550} & \textbf{600} & \textbf{650} & \textbf{700} & \textbf{750} & \textbf{800} & \textbf{850} & \textbf{900} & \textbf{950} & \textbf{1000} \\
    \midrule
    \textbf{-90\%} & 6\%   & 8\%   & 12\%  & 14\%  & 16\%  & 18\%  & 20\%  & 22\%  & 23\%  & 25\%  & 26\%  & 27\%  & 28\%  & 29\%  & 30\%  & 31\%  & 32\%  & 33\%  & 34\%  & 35\%  & 36\% \\
    \midrule
    \textbf{-80\%} & 8\%   & 12\%  & 17\%  & 20\%  & 24\%  & 26\%  & 29\%  & 31\%  & 33\%  & 35\%  & 36\%  & 38\%  & 40\%  & 41\%  & 42\%  & 44\%  & 45\%  & 46\%  & 47\%  & 49\%  & 50\% \\
    \midrule
    \textbf{-70\%} & 10\%  & 15\%  & 21\%  & 26\%  & 29\%  & 33\%  & 35\%  & 38\%  & 41\%  & 43\%  & 45\%  & 47\%  & 49\%  & 50\%  & 52\%  & 53\%  & 55\%  & 56\%  & 58\%  & 59\%  & 60\% \\
    \midrule
    \textbf{-60\%} & 12\%  & 18\%  & 25\%  & 30\%  & 34\%  & 38\%  & 42\%  & 45\%  & 47\%  & 50\%  & 52\%  & 54\%  & 56\%  & 58\%  & 60\%  & 61\%  & 63\%  & 64\%  & 66\%  & 67\%  & 68\% \\
    \midrule
    \textbf{-50\%} & 14\%  & 20\%  & 28\%  & 34\%  & 39\%  & 44\%  & 47\%  & 50\%  & 53\%  & 56\%  & 58\%  & 61\%  & 63\%  & 65\%  & 67\%  & 68\%  & 70\%  & 71\%  & 73\%  & 74\%  & 75\% \\
    \midrule
    \textbf{-40\%} & 16\%  & 23\%  & 32\%  & 39\%  & 44\%  & 49\%  & 53\%  & 56\%  & 59\%  & 62\%  & 64\%  & 67\%  & 69\%  & 71\%  & 73\%  & 74\%  & 76\%  & 77\%  & 78\%  & 80\%  & \cellcolor[rgb]{ .788,  1,  .882}81\% \\
\end{tabular}%
}
  \label{tab:tab19bis}%
\end{table}%

\begin{table}[H]
  \centering
  \caption{Skill percentage for a Sharpe ratio of 1.0 for \textbf{monthly} periods under various auto-correlation assumptions under Student assumptions (two-tailed test)}
  \vspace{0.25cm}
  \resizebox {0.8 \textwidth} {!} {
    \begin{tabular}{|r|r|r|r|r|r|r|r|r|r|r|r|r|r|r|r|r|r|r|r|r|}
    \toprule
    \multicolumn{1}{|p{5.5em}|}{\textbf{ \textbackslash{} Observ.\newline{}Correl.}} & \textbf{6} & \textbf{12} & \textbf{18} & \textbf{24} & \textbf{30} & \textbf{36} & \textbf{42} & \textbf{48} & \textbf{54} & \textbf{60} & \textbf{66} & \textbf{72} & \textbf{78} & \textbf{84} & \textbf{90} & \textbf{96} & \textbf{102} & \textbf{108} & \textbf{114} & \textbf{120} \\
    \midrule
    \textbf{-90\%} & 15\%  & 22\%  & 27\%  & 31\%  & 35\%  & 38\%  & 41\%  & 43\%  & 45\%  & 48\%  & 50\%  & 52\%  & 53\%  & 55\%  & 57\%  & 58\%  & 60\%  & 61\%  & 62\%  & 63\% \\
    \midrule
    \textbf{-80\%} & 20\%  & 29\%  & 36\%  & 41\%  & 45\%  & 49\%  & 53\%  & 56\%  & 58\%  & 61\%  & 63\%  & 65\%  & 67\%  & 69\%  & 71\%  & 72\%  & 74\%  & 75\%  & 76\%  & 78\% \\
    \midrule
    \textbf{-70\%} & 24\%  & 35\%  & 42\%  & 48\%  & 53\%  & 57\%  & 61\%  & 64\%  & 67\%  & 70\%  & 72\%  & 74\%  & 76\%  & 78\%  & 79\%  & \cellcolor[rgb]{ .788,  1,  .882}81\% & \cellcolor[rgb]{ .788,  1,  .882}82\% & \cellcolor[rgb]{ .788,  1,  .882}83\% & \cellcolor[rgb]{ .788,  1,  .882}85\% & \cellcolor[rgb]{ .788,  1,  .882}86\% \\
    \midrule
    \textbf{-60\%} & 28\%  & 40\%  & 48\%  & 55\%  & 60\%  & 64\%  & 68\%  & 71\%  & 74\%  & 77\%  & 79\%  & \cellcolor[rgb]{ .788,  1,  .882}81\% & \cellcolor[rgb]{ .788,  1,  .882}83\% & \cellcolor[rgb]{ .788,  1,  .882}84\% & \cellcolor[rgb]{ .788,  1,  .882}86\% & \cellcolor[rgb]{ .788,  1,  .882}87\% & \cellcolor[rgb]{ .788,  1,  .882}88\% & \cellcolor[rgb]{ .788,  1,  .882}89\% & \cellcolor[rgb]{ .788,  1,  .882}90\% & \cellcolor[rgb]{ .592,  1,  .776}91\% \\
\end{tabular}%
}
  \label{tab:tab20bis}%
\end{table}%

\subsubsection{Sharpe ratio of 1.5}

\vspace{-0.5cm}

\begin{table}[H]
  \centering
  \caption{Skill percentage for a Sharpe ratio of 1.5 for \textbf{daily} periods under various auto-correlation assumptions under Wald assumptions (two-tailed test)}
  \vspace{0.25cm}
  \resizebox {0.8 \textwidth} {!} {
    %
%
}
  \label{tab:tab21}%
\end{table}%

\begin{table}[H]
  \centering
  \caption{Skill percentage for a Sharpe ratio of 1.5 for \textbf{monthly} periods under various auto-correlation assumptions under Wald assumptions (two-tailed test)}
  \vspace{0.25cm}
  \resizebox {0.8 \textwidth} {!} {
    %
%
}
  \label{tab:tab22}%
\end{table}%

\begin{table}[H]
  \centering
  \caption{Skill percentage for a Sharpe ratio of 1.5 for \textbf{daily} periods under various auto-correlation assumptions under Student assumptions (one-tailed test)}
  \vspace{0.25cm}
  \resizebox {0.8 \textwidth} {!} {
    %
%
}
  \label{tab:tab23}%
\end{table}%

\begin{table}[H]
  \centering
  \caption{Skill percentage for a Sharpe ratio of 1.5 for \textbf{monthly} periods under various auto-correlation assumptions under Student assumptions (one-tailed test)}
  \vspace{0.25cm}
  \resizebox {0.8 \textwidth} {!} {
    %
%
}
  \label{tab:tab24}%
\end{table}%

\begin{table}[H]
  \centering
  \caption{Skill percentage for a Sharpe ratio of 1.5 for \textbf{daily} periods under various auto-correlation assumptions under Student assumptions (two-tailed test)}
  \vspace{0.25cm}
  \resizebox {0.8 \textwidth} {!} {
    %
%
}
  \label{tab:tab23bis}%
\end{table}%

\begin{table}[H]
  \centering
  \caption{Skill percentage for a Sharpe ratio of 1.5 for \textbf{monthly} periods under various auto-correlation assumptions under Student assumptions (two-tailed test)}
  \vspace{0.25cm}
  \resizebox {0.8 \textwidth} {!} {
    %
%
    }
  \label{tab:tab24bis}%
\end{table}%

\pagebreak
\subsubsection{Sharpe ratio of 2.0}
\vspace{-0.5cm}

\begin{table}[H]
  \centering
  \caption{Skill percentage for a Sharpe ratio of 2.0 for \textbf{daily} periods under various auto-correlation assumptions under Wald assumptions (two-tailed test)}
  \vspace{0.25cm}
  \resizebox {0.8 \textwidth} {!} {
    %
%
}
  \label{tab:tab25}%
\end{table}%

\begin{table}[H]
  \centering
  \caption{Skill percentage for a Sharpe ratio of 2.0 for \textbf{monthly} periods under various auto-correlation assumptions under Wald assumptions (two-tailed test)}
  \vspace{0.25cm}
  \resizebox {0.8 \textwidth} {!} {
    %
%
}
  \label{tab:tab26}%
\end{table}%

\begin{table}[H]
  \centering
  \caption{Skill percentage for a Sharpe ratio of 2.0 for \textbf{daily} periods under various auto-correlation assumptions under Student assumptions (one-tailed test)}
  \vspace{0.25cm}
  \resizebox {0.8 \textwidth} {!} {
    %
%
}
  \label{tab:tab27}%
\end{table}%

\begin{table}[H]
  \centering
  \caption{Skill percentage for a Sharpe ratio of 2.0 for \textbf{monthly} periods under various auto-correlation assumptions under Student assumptions (one-tailed test)}
  \vspace{0.25cm}
  \resizebox {0.8 \textwidth} {!} {
    %
%
}
  \label{tab:tab28bis}%
\end{table}%

\end{document}